\documentclass{article}

\usepackage{amssymb}
\usepackage{amsmath}

\usepackage{graphicx}

\newtheorem{theorem}{Theorem}[section]
\newtheorem{prop}[theorem]{Proposition}
\newtheorem{lem}[theorem]{Lemma}
\newtheorem{cor}[theorem]{Corollary}
\newtheorem{defi}[theorem]{Definition}
\newtheorem{remark}[theorem]{Remark}

\begin{document}
\title{Algebraic varieties in Birkhoff strata of the Grassmannian Gr$\mathrm{^{(2)}}$: Harrison cohomology and integrable systems}
\author{
B.G.Konopelchenko $^1$ and G.Ortenzi $^2$\\
$^1$ {\footnotesize Dipartimento di Fisica, Universit\`{a} del Salento }  \\
{\footnotesize and INFN, Sezione di Lecce, 73100 Lecce, Italy, \texttt{konopel@le.infn.it }} \\
 $^2$ {\footnotesize Dipartimento di Matematica Pura ed Applicazioni, } \\
{\footnotesize Universit\`{a} di Milano Bicocca, 20125 Milano, Italy, \texttt{giovanni.ortenzi@unimib.it} }
}
\maketitle
\begin{abstract}
Local properties of families of algebraic subsets $W_g$ in Birkhoff strata $\Sigma_{2g}$ of Gr$^{(2)}$ containing hyperelliptic curves of genus $g$ are studied. It is shown that the tangent spaces $T_g$ for $W_g$ are isomorphic to linear spaces of $2$-coboundaries. Particular subsets in $W_g$ are described by the intergrable dispersionless coupled KdV systems of hydrodynamical type defining a special class of $2$-cocycles and $2$-coboundaries in $T_g$. It is demonstrated that the blows-ups  of such $2$-cocycles and  
$2$-coboundaries and gradient catastrophes for associated integrable systems are interrelated.
\end{abstract}
\section{Introduction}
In this paper we continue the study of structure and properties of algebraic curves and varieties in Birkhoff strata $\Sigma_s$ of the Sato Grassmannian \cite{SS,SW,PS} within the approach proposed in \cite{KO}. It was shown in \cite{KO} that each strata $\Sigma_s$ contain subset $W$ closed with respect to pointwise multiplication which geometrically represents an infinite-dimensional algebraic variety defined by the relations
\begin{eqnarray}
&& p_jp_k-\sum_lC_{jk}^l p_l =0, \label{ppCp}\\
&& \sum_l \left( C_{jk}^lC_{lm}^r-C_{mk}^lC_{lj}^r \right)=0 ,\qquad j,k,m,r=0,1,2,3,\dots\ \label{CCuCC}.
\end{eqnarray}
Algebraically the relations (\ref{ppCp}) with the condition $C_{jk}^l=C_{kj}^l$ represent a table of multiplication of a commutative associative algebra $\mathcal{A}$ in the basis ($p_0,p_1,p_2,\dots$) while (\ref{CCuCC}) is the associativity condition of the structure constants $C^l_{jk}$.\par
In virtue of this algebro-geometrical duality, the subsets $W$ are of particular interest. Indeed, ``the associativity relations are so natural that any information deduced from them should have some kind of meaning'' (see \cite{Sha}, Ch. II, Sec. 1.3). On the other hand, due to its relation to commutative  algebras, the algebraic varieties defined by equations (\ref{ppCp},\ref{CCuCC}) are the natural objects for the Harrison's \cite{Hron} cohomology theory ``which is particularly applicable to the coordinate ring of algebraic varieties'' \cite{NR}.\par
Here we will study local properties of the subsets $W$ in Birkhoff strata of the Grassmannian Gr$^{(2)}$. This Grassmannian is an important specialization of the universal Sato Grassmannian Gr \cite{SS,SW,PS}. At the same time, all calculations are simplified drastically for Gr$^{(2)}$ that allows us to perform a complete analysis.\par
 In the previous work \cite{KO-p1} we showed that the Birkhoff strata $\Sigma_{2g}$ ($g=0,1,2,3,\dots$) in Gr$^{(2)}$ contain infinite-dimensional subsets $W_g$ defined by (\ref{ppCp},\ref{CCuCC}) with $p_{2n}=z^{2n}$, $n=0,1,2,3,\dots$ which are isomorphic to infinite families of coordinate rings for rational normal (Veronese) curves of all odd orders ($g=0$), for elliptic curves ($g=1$) and hyperelliptic curves ($g > 1$)
\begin{equation}
\label{Hpell}
 p_{2g+1}^2=\lambda^{2g+1}+\sum_{k=0}^{2g}u_k \lambda^k. 
\end{equation}
Hyperelliptic curve (\ref{Hpell}) at fixed $u_k$ is contained in a point of the subset $W_g$.
It was noted in \cite{KO} that tangent spaces for algebraic varieties in the Birkhoff strata of the Sato Grassmannian
 Gr are isomorphic to linear spaces of $2$-coboundaries. Here we will show that the tangent spaces for the varieties $W_g$ are isomorphic to the linear spaces of $2$-coboundaries of the special structure. In particular, $2$-cocycles $\psi_g(\alpha,\beta)$ and $2$-coboundaries $f_g(\gamma)$ for these varieties are related as
\begin{equation}
 \psi_g(p_{2k+1},p_{2k+1})=2p_{2k+1}f_g(p_{2k+1}), \qquad k=g,g+1,\dots\ .
\end{equation}
 We present also representations of the 2-cocycles and 2-coboundaries for $W_g$ in terms of the elements of the ideals $I(W_g)$ of these varieties.\par
We consider special  subvarieties $W_{gc}^I$ defined only by the equations (\ref{CCuCC}) for which the differential one-forms 
\begin{equation}
 \omega_c=\sum_{k=0}^g p_{2(g+k)+1}(z)dx_{2(g+k)+1}
\end{equation}
are closed.
It is shown that such varieties $W^I_{gc}$ are characterized by the closedness of only $2g+1$ one-forms. Moreover these subvarieties are described by $2g$ commuting $2g+1$ components hydrodynamical type systems. 
For the subvarieties $W_g^I$  of the subsets $W_g$ for which the one-forms 
\begin{equation}
\omega=\sum_{k=0}^\infty p_{2(g+k)+1}(z)dx_{2(g+k)+1} 
\end{equation}
are closed, the associated systems form infinite hierarchies of integrable systems. For $g=0$ it is the Burgers-Hopf hierarchy while for $g\geq 1$ they are dispersionless coupled Korteweg-de Vries (dcKdV) hierarchies. Solutions of these integrable hierarchies provide us a special class of 2-cocycles and 2-coboundaries.\par
At $g=1$ we consider also a particular class of the varieties $W_1$ for which $u_0=0$. It is shown that the corresponding reduced subset $W_{1c}$ represents a two-dimensional family of coordinate rings for the elliptic curve 
with the fixed point in the origin $(p_3,\lambda)=(0,0)$. The associated hierarchy is the dispersionless nonlinear Schroedinger (dNLS) equation or 1-layer Benney hierarchy. It is shown that for such hierarchy there exists  a simple tau-function $\phi$ such that the dNLS equation is equivalent to the Hirota-type equation 
\begin{equation}
\mathrm{det}\left(
 \begin{array}{cc}
  \phi_{x_3x_3} & \phi_{x_3x_5} \\
\phi_{x_5x_3} & \phi_{x_5x_5}
 \end{array}
\right)  + \left({\phi_{x_3x_3}}\right)^3=0.
\end{equation}
 Solutions of this equation provide the particular dNLS 2-cocycles and 2-coboundaries.\par
Interrelation between the properties of the dcKdV hierarchies describing the varieties $W_g$ and the corresponding 2-cocycles and 2-coboundaries is discussed too. It is observed that the blow-ups of the 2-cocycles and 2-coboundaries (i.e. an unbounded increase of their values) and the gradient catastrophe for the above hydrodynamical type system happen on the same subvarieties of finite codimension of the affine space of the variables $x_k$.\par
The paper is organized as follows. Birkhoff strata in the Grassmannian and the results of the paper \cite{KO-p1} are briefly described in section \ref{sec-res}. Harrison cohomology of the varieties $W_{gc}$ and $W_g$ are discussed in section 3. Big cell in Gr$^{(2)}$ and associated Burgers-Hopf hierarchy are considered in section 4. Section 5 is devoted to the stratum $\Sigma_2$ which contains family of elliptic curves and corresponding integrable systems. Deformations of moduli $g_2$ and $g_3$ of the elliptic curve and associated 2-cocycles are studied in section 6. Deformations of elliptic curve with a fixed point in the origin described by the dispersionless NLS equation and associated Hirota type equation are considered in section 7. Hyperelliptic curves in $W_g$ and corresponding dispersionless coupled KdV (dcKdV) hierarchies are discussed in section 8. Interpretation of the ideals of varieties $W^I_g$ as the Poisson ideals is presented in section 9. Interrelation between cohomology blow-ups and gradient catastrophe for dcKdV hierarchies is discussed briefly in section 10. Comparison with other approaches is given in section 11.
\section{Birkhoff strata in the Grassmannian Gr$^{(2)}$ and associated algebraic varieties}
\label{sec-res}
For the completeness we recall here briefly the basic facts about Birkhoff strata in Gr$^{(2)}$ and corresponding results of the paper \cite{KO-p1}. \par
Sato Grassmannian Gr can be viewed as the set of closed vector subspaces in the infinite dimensional set of all formal Laurent series with coefficients in $\mathbb{C}$ with certain special properties (see e.g. \cite{SW,PS}). Each subspace W$\in$Gr possesses an algebraic basis $(w_0(z),w_1(z),w_2(z),\dots)$ with the basis elements
\begin{equation} 
\label{wn}
 w_n=\sum_{k=-\infty}^n a_k^n z^k
\end{equation}
 of finite order $n$. Grassmannian Gr is a connected Banach manifold which exhibits a stratified structure \cite{SW,PS}, i.e. Gr=$\bigcup_{S} \Sigma_S$ where the stratum $\Sigma_S$ is a subset in Gr formed by elements of the form (\ref{wn}) such that possible values $n$ are given by the infinite set $S=\{s_0,s_1,s_2,\dots\}$ of integers $s_n$ with $s_0<s_1<s_2<\dots $ and $s_n = n$ for large $n$. Big cell $\Sigma_0$ corresponds to $S=\{0,1,2,\dots \}$. Other strata are associated with the sets $S$ different from $S_0$. \par
Gr$^{(2)}$ is the subset of elements $W$ of Gr obeying the condition $z^2 \cdot W \subset W$ \cite{SW,PS}. This condition imposes strong constraints on the Laurent series and on the structure of the strata. Namely, Birkhoff stratum $\Sigma_S$ in Gr$^{(2)}$ corresponds to the sets $S$ such that $S+2 \subset S$, i.e. all possible $S$ having the form \cite{SW,PS}
\begin{equation}
\label{Sm} 
S_m=\{-m,-m+2,-m+4,\dots,m,m+1,m+2,\dots \}
\end{equation}
with $m=0,1,2,\dots$. Codimension of $\Sigma_m$ is $m(m+1)/2$. One has Gr$^{(2)}=\bigcup_{m \geq 0} \Sigma_m$. \par
In this paper we will consider only strata $\Sigma_{2g}$, $g=0,1,2,\dots$.
Stratum $\Sigma_{2g}$ with arbitrary $g$ is characterized by $S=\{ -2g, -2g+2, -2g+4, \dots, 0, 2, 4, \dots, 2g, 2g+1, 2g+2, \dots \}$. So it does not contain, in particular, $g$ elements of the orders $ 1, 3, 5, \dots, 2g-1$ and the positive order elements of the canonical basis are given by
\begin{equation}
 \label{series-S2n}
\begin{split}
 p_0=& 1+\sum_{k \geq 1} \frac{H^0_k}{z^k},\\
 p_j=& z^j+\sum_{k=0}^{j-1}H^j_{-2k-1}z^{2k+1} + \sum_{k \geq 1} \frac{H^j_k}{z^k}, \qquad j=2,4,6,\dots, 2g-2, \\
 p_j=& z^j+\sum_{k=0}^{g-1}H^j_{-2k-1}z^{2k+1} + \sum_{k \geq 1} \frac{H^j_k}{z^k} \qquad j=2g,2g+1,2g+2,2g+3,\dots\ .
\end{split}
\end{equation}
where $H^j_k$ are arbitrary complex values parameters. \par
It was shown in \cite{KO} and \cite{KO-p1} that
\begin{prop}
The stratum $\Sigma_{2g}$ for $g=1,2,3,\dots$ contains maximal subset $W_{g}$ closed with respect to pointwise multiplication with elements of the form 
\begin{equation}
 w=\sum_{k \geq 0} a_{2k} \lambda^k+ \sum_{k \geq g} b_{2k+1}(\lambda) p_{2k+1}
\end{equation}
with parameters $H^j_k$ obeying the constraints 
\begin{equation}
 \label{Heven-S2n}
\begin{split}
 H^{2m}_k =& 0, \qquad m =0,1,2, \dots, \ k= -2g+2, -2g+4, \dots, -2, 0, 1, 2, 3, \dots, \\
H^{2m+1}_{2k} =& 0, \qquad m = 0,1,2, \dots \ k = -g , -g+1, -g+2, \dots 
\end{split}
\end{equation}
and
\begin{equation}
\label{H-S2n}
\begin{split}
&H^{2j+1}_{2(l+k)+1}-H^{2(j+k)+1}_{2l+1}-\sum_{s=-g}^{k-1}H^{2j+1}_{2s+1}H^{2(k-s)-1}_{2l+1}=0, \\
&H^{2j+1}_{2(l+k)+1}+H^{2k+1}_{2(l+j)+1}
+\sum_{s=-g}^{-1} H^{2j+1}_{2s+1}H^{2k+1}_{2(l-s)-1}
+\sum_{r=-g}^{-1} H^{2k+1}_{2r+1}H^{2j+1}_{2(l-r)-1}
+\sum_{s=0}^{l-g} H^{2j+1}_{2s+1}H^{2k+1}_{2(l-s)-1}=0.
\end{split}
\end{equation} 
\end{prop}
and $p_{2m}=\lambda^m$, $m=0,1,2,\dots$, $\lambda=z^2$. One also has 
\begin{equation}
\label{C2g+1}
 C_{2g+1}\equiv p_{2g+1}^2-\left( \lambda^{2g+1}+\sum_{k=0}^{2g} u_k \lambda^k\right)=0
\end{equation}
where $u_k$ are certain polynomials in $H^j_k$ and
\begin{equation}
\label{lg}
 l^{(g)}_{2m+1}\equiv p_{2m+1}-\alpha_m(\lambda)p_{2g+1}=0, \qquad m=g+1,g+2,\dots
\end{equation}
where $\alpha_m(\lambda)$ are polynomials in $\lambda$. \par 
The subsets $W_g$ have the following algebraic and geometrical interpretation
\begin{prop}
 \label{prop-Wg}
\cite{KO-p1}
Each point of the subset $W_g$ ($g$ $=$ $0,$ $1,$ $2, \dots$) corresponding to  fixed values of $H^j_k$ which obey the conditions (\ref{Heven-S2n}), (\ref{H-S2n}) is an infinite dimensional commutative algebra $A_g$  isomorphic
to $\mathcal{C}[\lambda,p_{2g+1}]/ C_{2g+1}$.
The subset $W_g$ is isomorphic to an infinite 
dimensional algebraic variety  defined by the equations
\begin{equation}
\label{fjk}
 f_{jk}\equiv p_j p_k -C_{jk}^l p_l =0 
\end{equation}
 and by the relations (\ref{Heven-S2n}) and (\ref{H-S2n}) in the affine space with local coordinates
\begin{equation}
 \lambda, p_{2(g+m)+1}, H^{2(g+m)+1}_{-(2g-k)-1}, \qquad k,m=1,2,3,\dots\ .
\end{equation}
 Its ideal is 
\begin{equation}
\label{IGamma}
 I_g=\langle C_{2g+1}, l^{(g)}_{2g+3},l^{(g)}_{2g+5},\dots \rangle
\end{equation}
where $l^{(g)}_{2m+1}$ are given by (\ref{lg}).
\end{prop}
With small abuse of notation we will refer to this infinite dimensional algebraic variety as $W_g$ in the 
rest of the paper. \par
The relations (\ref{Heven-S2n}) and (\ref{H-S2n}) are equivalent to the associativity conditions
\begin{equation}
\label{CCCC}
 \sum_s C_{jk}^s C_{ls}^r- C_{lk}^s C_{js}^r=0
\end{equation}
for the structure constants $C_{jk}^l$ given by
\begin{equation}
\begin{split}
 C_{2j,2k}^{2l}=&\delta^l_{j+k}, \\
C_{2j+1,2k}^{2l+1}=&\delta^l_{j+k} + H^{2j+1}_{2(k-l)-1},\\
C_{2j,2k}^{2l}=&\delta^l_{j+k}
+H^{2j+1}_{2(j-l)+1}
+H^{2k+1}_{2(k-l)+1} 
+\sum_{s=-g-1}^{-1}\sum_{r=-g-1}^{-1}H^{2j+1}_{2s+1}H^{2k+1}_{2r+1}\delta^l_{-2(s+r+1)}\\&
+\sum_{s=-g-1}^{-1}\sum_{r=0}^{-s-1}H^{2j+1}_{2s+1}H^{2k+1}_{2r+1}\delta^l_{-2(s+r+1)}
+\sum_{r=-g-1}^{-1}\sum_{s=0}^{-r-1}H^{2j+1}_{2s+1}H^{2k+1}_{2r+1}\delta^l_{-(s+r+1)}.  
\end{split}
\end{equation}
The variety  $W_g$ is the intersection of quadrics (\ref{fjk}) and (\ref{CCCC}). \par
Sections of these varieties $W_g$ by the planes with all $H^j_k=const$ represents well known algebraic curves.
At $g=0$ the relations (\ref{C2g+1}) and (\ref{lg}) are equivalent to the following
\begin{equation}
\label{p-g0}
 \begin{split}
  \lambda=&p_1^2-2H^1_1, \\
p_3=& p_1^3-3H^1_1 p_1, \\
p_5=& p_1^5-5H^1_1 p_1^3+\frac{15}{2} {H^1_1}^2p_1,\\
\dots \ .&
 \end{split}
\end{equation}
The formulae (\ref{p-g0}) represent the well known parameterization of the rational normal (Veronese curves)
of odd order (see e.g. \cite{Har}). So, for the big cell the subset $W_0$ is the infinite dimensional family of the infinite tower  of the Veronese curves of all odd orders. \par
For $g=1$ the relations (\ref{C2g+1}) and (\ref{lg}) take the form
\begin{eqnarray}
 \label{C3}
C_3&=&p_3^2-\left(\lambda^3+u_2\lambda^2+u_1\lambda+u_0\right), \\
\label{l1}
l^{(1)}_{2m+1}&=&p_{2m+1} -\left( \lambda^{m-1} -\sum_{k=0}^{m-2} H^{2(m-k)-1}_{-1}\lambda^k \right)p_3, \qquad m=2,3,\dots .
\end{eqnarray}
Thus, the subset $W_1$ is the infinite family of the coordinate ring for the family of elliptic curves (\ref{C3}).\par
At $g>1$ one has a hyperelliptic curve of genus $g$ defined by the equation (\ref{C2g+1}) and an infinite family of its coordinate rings.\par
We would like to note that one can view these subsets $W_g$ in different ways. First one can interpret them as the infinite families of the deformed basic curves  (rational normal curves ($g=0$), elliptic ($g=1$) or hyperelliptic ($g>1$) curves) parameterized by the variables $H^j_k$ obeying the constraints (\ref{C2g+1}) and (\ref{lg}) (or (\ref{Heven-S2n}) and (\ref{H-S2n})). Second since these subsets $W_g$ are defined by the quadratic equations (\ref{fjk}) and (\ref{CCCC}) they are isomophic infinite dimensional algebraic varieties in the affine spaces with the local coordinates $\lambda,p_{2g+1},p_{2g+3},\dots,H^{j}_k$. Finally for all $p_j=const$ and $\lambda=const$ the subsets $W_g$ coincide with the varieties of structure constants  $C^l_{jk}$ of commutative associative algebras which are of their own interest (see e.g. \cite{Sha}). We will refer to such subvarieties as $W_{gc}$. All these aspects of the subsets $W_g$ are equally important and arise in various applications. 
\section{Harrison cohomology}
\label{sec-Har} 
Study of the properties of the varieties $W_g$ described in the previous section begins, as usually, with the analysis of their local structure. A standard way to do this is to consider a tangent bundle $T_{W_g}$ for $W_g$. In virtue of the equations (\ref{ppCp}) and (\ref{CCuCC}), $T_{W_g}$ is defined by the following systems of linear equations
\begin{eqnarray}
 &&\pi_jp_k+p_j\pi_k-\sum_l\Delta_{jk}^lp_l-\sum_l C_{jk}^l\pi_l=0 \label{Talg}, \\
&&\sum_l \left(\Delta_{jk}^lC_{lm}^p+C_{jk}^l\Delta_{lm}^p-\Delta_{mk}^lC_{lj}^p-C_{mk}^l\Delta_{lj}^p\right)=0, \qquad j,k,m,p \in S_{1,2,\dots,n}. \label{Tass}
\end{eqnarray}
where $\pi_j,\Delta_{jk}^l \in T_{W_g}$. The subsystem (\ref{Tass}) defines a tangent bundle for the subvarieties $W_{gc}$. The linear system
\begin{eqnarray}
 &&\pi_j^*p_k+p_j\pi_k^*-\sum_l\Delta^{*l}_{jk}p_l-\sum_l C_{jk}^l\pi_l^*=0 \label{cTalg}, \\
&&\sum_l \left(\Delta_{jk}^{*l}C_{lm}^p+C_{jk}^l\Delta_{lm}^{*p}-\Delta_{mk}^{*l}C_{lj}^p-C_{mk}^l\Delta_{lj}^{*p}\right)=0, \qquad j,k,m,p \in S_{1,2,\dots,n}. \label{cTass}
\end{eqnarray}
defines a cotangent bundle $T^*_{W_g}$ for the variety $W_{g}$.
Elements of $T_{W_g}$ and  $T_{W_g}^*$ can be realized, as usual (see e.g. \cite{Sha,Har,Arn}), as
\begin{eqnarray}
 \pi_j&=& X (p_j), \qquad \Delta^l_{jk}=X(C^l_{jk}), \\
 {\pi^*}_j&=& d p_j, \qquad {\Delta^*}^l_{jk}=d C^l_{jk} ,
\end{eqnarray}
where $X$ is a vector field on $W_g$ and $d p_j, \, d C^l_{jk}$ are differentials of $ p_j, \, C^l_{jk}$. In particular, the system (\ref{Tass}) for $d C^l_{jk}$ defines a module of differential for the variety $W_{gc}$ of the structure constants $C_{jk}^l$.\par
Cohomology theory of commutative associative algebras proposed by Harrison in \cite{Hron} is the most appropriate to analyze local properties of the varieties $W_{gc}$. \par
Harrison's cohomology is a commutative version of the classical Hochschild cohomology for associative algebras \cite{Hoc,Hoc1}. For self-consistence we remind it briefly. Let $K$ be a field and $A$ be an associative algebra (possibly infinite dimensional) over $K$. Let $\Lambda$ be a two-sided $A$-module. The set of all n-dimensional $\Lambda$-cochains (or $(n,\Lambda)$-cochains) of $A$ and is denoted by $C^n(A,\Lambda)$. The coboundary operator $\delta$ maps each $C^n(A,\Lambda)$ into $C^{n+1}(A,\Lambda)$ according to the rule: For $f\in C^n(A,\Lambda)$ and $a_1,\dots,a_n \in A$ 
\begin{equation}
\begin{split}
 (\delta f)(a_1,a_2,\dots,a_{n+1})=&a_1 \cdot f(a_2,\dots,a_{n+1}) +\sum_{k=1}^n (-1)^k f(a_1,\dots,a_k\cdot a_{k+1},\dots,a_{n+1})\\ & +(-1)^{n+1} f(a_1,\dots,a_{n})\cdot a_{n+1}
\end{split}
\end{equation}
where the dot denotes the product in $A$. For any cochain $f$ one has $\delta(\delta f)=0$. An element $f\in C^n(A,\Lambda)$ is called an n-dimensional cocycle if $\delta f=0 $. An element of the form $\delta g$ with $g\in C^{n-1}(A,\Lambda)$ is called an n-dimensional
coboundary. For instance, 2-cocycle is a bilinear map $f$ from $A \times A$ to $\Lambda$ such that  
\begin{equation}
\label{eqn-cocy-gen}
 a_1 \cdot f(a_2,a_3) - f(a_1 \cdot a_2,a_3) +  f(a_1, a_2 \cdot a_3)- f(a_1,a_2)\cdot a_{3}=0
\end{equation}
while 2-coboundary $\phi$ is defined by 
\begin{equation}
\label{eqn-cobo-gen}
 \phi(a_1,a_2)=a_1 \cdot g(a_2)-g(a_1 \cdot a_2) +g(a_1) \cdot a_2
\end{equation}
where $g$ is a linear mapping.\par
The coboundaries form a subgroup $B^n(A,A)$ of the group $Z^n(A,\lambda)$ of n-dimensional cocycles. The n-dimensional cohomology group $H^n(A,\Lambda)$ is defined as the quotient  $Z^n(A,\lambda)/ B^n(A,\lambda)$ \cite{Ger1,Ger2}.\par
In the case of commutative associative algebras considered by Harrison, one has some new properties, but on the other hand, one can define only first three cohomology moduli $H^1,H^2,H^3$ \cite{Hron}. Main advantage of the Harrison cohomology theory is that it is quite suitable for an analysis of local properties of algebraic varieties \cite{Hron,NR,Ger1,Ger2,LS,ScSt,Fro}.\par
Particular examples of algebraic varieties defined by the associativity condition for structure constants of commutative associative alegbras in a fixed basis $\{p_j\}$ are of great general interest (see \cite{Sha}, Ch. II, Sec 1.3). Tangent spaces for such algebras are defined by equation (\ref{Talg},\ref{Tass}). Introducing the bilinear mapping  $\psi_g(\alpha,\beta)$ to a tangent space of $A_g$ defined by (see \cite{Sha}, p.91)
\begin{equation}
 \label{2-cocy}
 \psi_g(p_j,p_k)=\sum_l \Delta^l_{jk} p_l,
\end{equation}
 one observes that equations (\ref{Talg}) are equivalent to
\begin{equation}
 p_j\psi_g(p_k,p_l) -\psi_g(p_jp_k,p_l)+\psi_g(p_j,p_kp_l)-p_l\psi_g(p_j,p_k)=0 \qquad j,k,l =2g+l,\dots
\end{equation}
or to
\begin{equation}
\label{eqn-cocy}
 \alpha \psi_g (\beta,\gamma)- \psi_g (\alpha\beta,\gamma)+ \psi_g (\alpha,\beta\gamma)
-\gamma \psi_g (\alpha,\beta)=0
\end{equation}
where $\alpha,\beta,\gamma\in W_{gc}$. \par
 So comparing (\ref{eqn-cocy}) and (\ref{eqn-cocy-gen}), one concludes that the tangent spaces in a point of the varieties $W_{gc}$ of the structure constants are isomorphic to linear spaces
 of 2-cocycles on $W_{gc}$ \cite{Sha}. \par 
These 2-cocycles exhibit more specific property being considered in larger varieties $W_g$. Indeed, one observes that
the system (\ref{Talg}) is equivalent to the equation
\begin{equation}
 \label{eqn-cob}
\psi_g(\alpha,\beta)=\alpha f_g(\beta)+\beta f_g(\alpha)-f_g(\alpha \beta)
\end{equation}
 where $f_g(\alpha)$ denotes linear map defined by the relations $f_g(p_k)=\pi_k$. So,
\begin{equation}
  \label{eqn-psi-f}
\psi_g(\alpha,\beta)=(\delta f_g)(\alpha,\beta).
\end{equation}
 Thus, due to the definition (\ref{eqn-cobo-gen}), $\psi_g(\alpha,\beta)$ is a $2$-coboundary and one has
\begin{prop}
Tangent spaces of the varieties $W_g$ are isomorphic to the linear spaces of 2- coboundaries (see also \cite{KO}).
\end{prop}
Consequently the Harrison's cohomology moduli $H^2(W_g)$ and $H^3(W_g)$ vanish.\par
In more general setting the system (\ref{Talg}) and (\ref{Tass}) defines any $W_g$-module $E$. So, for the same reason as above, a $W_g$-module $E$ and, in particular, the cotangent spaces of $W_g$ are isomorphic to linear spaces of 2-coboundaries. \par
Due to the constraint $z^2 W_g \subset W_g$, i.e. $p_{2j}=z^{2j}$, the two cocycles  $\psi_g(\alpha,\beta)$ and linear maps $f_g(\alpha)$ have certain specific properties. Indeed, taking into account the explicit form of equations (\ref{ppCp}) for the varieties $W_g$ \cite{KO-p1}, i.e. ($p_{2j}=z^{2j}$)
\begin{equation}
\label{ppCp-exp}
\begin{split} 
 p_{2j} p_{2k} =& p_{2(j+k)}, \\
 p_{2j} p_{2k+1} =& p_{2(j+k)+1} +\sum_{s=-g}^{k-1}H^{2j+1}_{2s+1}p_{2(k-s)-1},\\
p_{2j+1} p_{2k+1} =& p_{2(j+k+1)}
+\sum_{s=-g}^j H^{2j+1}_{2s+1}p_{2(j-s)}
+\sum_{s=-g}^k H^{2k+1}_{2s+1}p_{2(k-s)} \\
&+\sum_{s=-g}^{-1}\sum_{r=-g}^{-1}H^{2j+1}_{2s+1}H^{2k+1}_{2r+1}p_{-2(s+r+1)}
 +\sum_{s=-g}^{-1}\sum_{r=0}^{-s-1}H^{2j+1}_{2s+1}H^{2k+1}_{2r+1}p_{-2(s+r+1)}\\
&+\sum_{r=-g}^{-1}\sum_{s=0}^{-r-1}H^{2j+1}_{2s+1}H^{2k+1}_{2r+1}p_{-2(s+r+1)},
\end{split}
\end{equation}
and formulae (\ref{Talg}), (\ref{Tass}), one finds that
\begin{equation}
 \label{psip0pk-exp}
\psi_g(p_0,p_k)=0, \qquad k=2g+1,2g+3,\dots,
\end{equation}
\begin{equation}
 \label{psip2np2m-exp}
\psi_g(p_{2n},p_{2m})=0,\quad f_g(p_{2n})=0 \qquad n,m=0,1,2,\dots,
\end{equation}
and
\begin{equation}
 \label{psipdpd-exp}
\psi_g(p_{2k+1},p_{2k+1})= 2p_{2k+1}f_g(p_{2k+1}),\quad f_g(p_{2n})=0 \qquad k=g,g+1,g+2,\dots \ .
\end{equation}
 Relation (\ref{psipdpd-exp}) between 2-cocycles and 2-coboundaries allows us to find explicit form of the map $f_g(\alpha)$. Using (\ref{ppCp-exp}), one first finds that
\begin{equation}
 \label{p2k+1p2k+1L}
p_{2k+1}^2=\lambda^{2k+1}+\sum_{m=0}^{2k} v_m \lambda^m
\end{equation}
where $v_{k}$ are certain polynomials of $H^j_k$. Hence   
\begin{equation}
 \psi_g(p_{2k+1},p_{2k+1}) = \sum_{m=0}^{2k} \Delta_m z^{2m}
\end{equation}
where $\Delta_m \in T_{W_g}$. So one finds that
\begin{equation}
 f_g(p_{2k+1})=\frac{\sum_{m=0}^{2k} \Delta_m z^{2m}}{2p_{2k+1}},\qquad k=g,g+1,g+2,\dots\ .
\end{equation}
Since 
\begin{equation}
\label{p2k+1H}
 p_{2k+1}=z^{2k+1} +\sum_{m=-g}^\infty \frac{H^{2k+1}_{2m+1}}{z^{2m+1}}
\end{equation}
and $H^j_k$ obey the associativity conditions (\ref{H-S2n}), $f(p_{2k+1})$ have the form 
\begin{equation}
  f_g(p_{2k+1})=\sum_{m=-g}^\infty \frac{A^{2k+1}_{2m+1}}{z^{2m+1}}.
\end{equation}
 Then  a straightforward calculation show that 
\begin{equation}
 {A^{2k+1}_{2m+1}}={\Delta^{2k+1}_{2m+1}}
\end{equation}
where ${\Delta^{2k+1}_{2m+1}} \in T_{W_g}$.\par
Thus one has
\begin{equation}
 f_g(p_{2k+1})=\sum_{m=-g}^\infty \frac{\Delta^{2k+1}_{2m+1}}{z^{2m+1}} =:\Delta(p_{2k+1})
\end{equation}
that is a quite natural result if one treats $p_{2k+1}$  as a Laurent series (\ref{p2k+1H}). Instead the relation
(\ref{psipdpd-exp}) seems to be far from trivial.\par
Finally we note that that considering a vector field $X_c$ acting in the variety $W_{gc}$, one can get a simple realization of the mapping $\psi_g(\alpha,\beta)$ and $f_g(\alpha)$ namely
\begin{equation}
 \label{psif-map}
f(p_j)=X_c ( p_j), \qquad \psi(p_k,p_l)=-X_c ( f_{kl}).
\end{equation}
\section{Integrable subvariety of $\mathbf{W_0}$ in the big cell and Burgers-Hopf hierarchy}
\label{sec-int-0} 
A way to select particular classes of the varieties $W_g$ or their subvarieties is to require that  their tangent or cotangent bundles have some additional properties. Cotangent bundles $\Omega_{W_g}$ for the variety $W_g$ is isomorphic to module $\Omega_g$ of differentials over $W_g$ defined by the relations (\ref{Talg}),(\ref{Tass}). 
Having in mind the concept of Lagrangian submanifolds of symplectic manifolds (see e.g. \cite{Arn}) it is quite natural for this purpose to consider special subsets in $W_g$ for which certain one-forms are closed.\par
Let us begin with the simplest case of the big cell. For $W_0$ one has \cite{KO-p1} 
\begin{equation}
\label{Cjkl-S0}
 \begin{split}
 C_{2n,2m}^{2l}&=\delta_{m+n}^l, \\
  C_{2n,2m+1}^{2l+1}&=\delta_{{m+n}}^l+ H^{2m+1}_{2(n-l)-1}, \\
C_{2n+1, 2m+1}^{2l}&= \delta_{{m+n+1}}^l+H^{2n+1}_{2(m-l)+1}+H^{2m+1}_{2(n-l)+1},
 \end{split}
\end{equation}
and the subset $W_{c0}$ is defined by the equation
\begin{equation} 
\label{S0-H}
\begin{split}
&H^{2m+1}_{2(k+n)+1}-H^{{2(m+n)+1}}_{2k+1}-\sum_{s=0}^{n-1} H^{2m+1}_{2s+1}H^{2(n-s)-1}_{2k+1}=0,\\
&H^{2m+1}_{2(k+n)+1}+H^{2n+1}_{2(k+m)+1}+\sum_{l=0}^{k-1} H^{2m+1}_{2l+1}H^{2n+1}_{2(k-l)-1}=0.
\end{split}
\end{equation}
Relations (\ref{S0-H}) imply that 
\begin{equation}
\label{symm-S0}
 kH^j_k=jH_j^k, \dots j,k=1,2,3,\dots\ .
\end{equation}
The lowest of the relations (\ref{S0-H}) and (\ref{symm-S0}) are
\begin{equation}
 \begin{split}
 & \left({H^1_{{1}}}\right)^{2}+2\,H^1_{{3}}=0,\\
&H^1_{{5}}+H^1_{{3}}H^1_{{1}}=0,\\
&H^3_3-3\,H^1_{{5}}-3\,H^1_{{3}}H^1_{{1}}-\left({H^1_{{1}}}\right)^{3}=0,\\
&2\,H^1_{{7}}+2\,H^1_{{5}}H^1_{{1}}+{H^1_{{3}}}^{2}=0,\\
&\dots
 \end{split}
\end{equation}
and
\begin{equation}
 3H^1_3=H^3_1, \quad 5H^1_3=H^5_1, \quad 7H^1_3=H^7_1, \quad \dots\ .
\end{equation}
These equations are equivalent to
\begin{equation}
\label{Hk111-S0}
 \begin{split}
  H^3_1+ \frac{3}{2}\left(H^1_1\right)^2&=0,\\
  H^5_1- \frac{5}{2}\left(H^1_1\right)^3&=0,\\ 
  H^3_3-\left({H^1_{{1}}}\right)^{3}&=0,\\
  H^7_1+ \frac{35}{8}\left(H^1_1\right)^4&=0,\\
\dots& \ .
 \end{split}
\end{equation}
In general one has
\begin{equation}
 H^{2k-1}_1- 2^k (2k-1) \binom{1/2}{k} \left(H^1_1\right)^k=0, \qquad k=1,2,3,\dots 
\end{equation}
and
\begin{equation}
 H^{2j+1}_{2k+1}+\alpha_{jk} \left(H^1_1\right)^{j+k+1}=0, \qquad k=1,2,3,\dots 
\end{equation}
where $\alpha_{jk}$ are some constants. So all $H^j_k$ are parameterized by the single variable $H^1_1$. Hence $W_{0c}$ is the one-dimensional variety and $W_0$ is a one parametric family of the towers of rational normal curves of all odd orders.\par
Thus the situation in the Grassmannian  Gr$^{(2)}$ is drastically degenerate with respect to the general Sato Grassmannian  where in the variety $W_0$ in the big cell one has the infinite-parametric family of rational normal curves \cite{KO}. Having in mind this degeneration we will proceed in way which will be applicable to other strata in Gr$^{(2)}$ and in the general Grassmannian.\par
Thus, let $x_1$ and $x_3$ are local coordinate in variety $W_{0c}$. Let us consider a subvariety $W_{0c}^I$ 
for which the one-form
\begin{equation}
\label{w31-S0}
 \omega_{31}=H^3_1dx_3+H^1_1dx_1,
\end{equation}
is closed. So in $W_{0c}^I$ on has 
\begin{equation}
 \label{dH3113-S0}
\frac{\partial H^3_1}{ \partial x_1} - \frac{\partial H^1_1}{ \partial x_3}=0. 
\end{equation}
Combining equations (\ref{Hk111-S0}) and (\ref{dH3113-S0}), one gets 
\begin{equation}
 \label{BH-S0}
\frac{\partial H^1_1}{ \partial x_3} +3H^1_1 \frac{\partial H^1_1}{ \partial x_1}=0,
\end{equation}
that is the well known Burgers-Hopf (BH) equation. On the other hand condition   (\ref{dH3113-S0}) locally implies the existence of a function $S_1$ such that
\begin{equation}
 H^3_1=\frac{\partial S_1}{ \partial x_3}, \qquad H^1_1=\frac{\partial S_1}{ \partial x_1}
\end{equation}
and 
\begin{equation}
 \omega_{31}=dS_1.
\end{equation}
Here and in the rest of the paper our consideration is pure local.
Substitution of the second of equations (\ref{Hk111-S0}) in the first one gives
\begin{equation}
 \frac{\partial S_1}{ \partial x_3} +\frac{3}{2} \left(\frac{\partial S_1}{ \partial x_1}\right)^2=0
\end{equation}
that is the potential form of the BH equation (\ref{BH-S0}).\par
Using the relations (\ref{Hk111-S0}) , one gets the following 
\begin{lem}
 Equation (\ref{dH3113-S0}) implies that 
\begin{equation}
 \frac{\partial H^3_k}{ \partial x_1} - \frac{\partial H^1_k}{ \partial x_3}=0 
\end{equation}
for all $k=1,3,5,7,\dots$ and, hence, 
\begin{equation}
 \frac{\partial p_3(z)}{ \partial x_1} - \frac{\partial p_1(z)}{ \partial x_3}=0.
\end{equation}
\end{lem}
Thus the closedness of the form (\ref{w31-S0}) implies the closedness of the infinite set of one-forms given by
\begin{equation} 
\omega_{31}(z)=p_3(z)dx_3+p_1(z)dx_1= z^3dx_3+zdx_1+\sum_{k\geq 1}^{\infty} \frac{1}{z^{2k+1}} (H^3_{2k+1}dx_3+H^1_{2k+1}dx_1).
\end{equation}
One can easily repeat such a construction starting with any closed one-form 
\begin{equation}
 \omega_{jk}=H^j_1dx_j+H^k_1dx_k,
\end{equation}
instead of the form (\ref{w31-S0}) and obtaining the infinite set of the forms
\begin{equation}
 \omega_{jk}(z)=p_j(z)dx_j+p_k(z)dx_k ,
\end{equation}
where $j,k$ are any two distinct indices of the set $\{1,3,5,\dots \}$.
Combining these results, one obtains
\begin{prop}
\label{w-S0-prop}
 Existence of the closed one-forms 
\begin{equation}
\label{wjk-S0}
 \omega_{jk}=H^j_1dx_j+H^k_1dx_k,
\end{equation} 
for any two distinct indices $j,k$ of the set $\{1,3,5,\dots \}$, is a necessary and sufficient condition for the closedness of the one-form
\begin{equation}
\label{w-S0}
 \omega=\sum_{j=1}^\infty p_j(z)dx_j.
\end{equation}
\end{prop}
The closedness of the form (\ref{w-S0}) locally implies the existence of the formal Laurent series (action)
\begin{equation}
 \label{S-S0}
S(z,x)=\sum_{k=0}^\infty z^{2k+1}x_{2k+1}+\sum_{m=0}^\infty \frac{S_{2m+1}}{z^{2m+1}}
\end{equation}
such that
\begin{equation}
 p_j=\frac{\partial S}{\partial x_j}, \qquad j=1,3,5,\dots\ .
\end{equation}
In particular 
\begin{equation}
\label{Hjk-S0}
 H^j_k=\frac{\partial S_k}{\partial x_j}.
\end{equation}
Substitution of the expressions (\ref{Hjk-S0}) into algebraic relations (\ref{Hk111-S0}), gives an infinite set of differential equations which is equivalent to the following
\begin{equation}
\label{BH-hier}
 \frac{\partial H^1_1}{\partial x_{2k-1}}- 2^k k(2k-1) \binom{1/2}{k} \left(H^1_1\right)^{k-1}\frac{\partial H^1_1}{\partial x_1}=0, \qquad k=1,2,3,\dots\ . 
\end{equation}
This is a standard form of the BH hierarchy. Another form of the BH hierarchy is given by the infinite system of the Hamilton-Jacobi type equations
\begin{equation}
\label{HamJac-S0}
 \frac{\partial S}{\partial x_j}\frac{\partial S}{\partial x_k}-\sum_{l=1,3,5,\dots}C_{jk}^l\frac{\partial S}{\partial x_l}=0
\end{equation}
 with $S$ given by (\ref{S-S0}) and $C_{jk}^l$ given by (\ref{Cjkl-S0}).\par
Closed one-forms of type  (\ref{w-S0}) and action $S(x,z)$ (\ref{S-S0}) are basis objects in various approaches to the dispersionless integrable hierarchies \cite{Kod,TT,TT2}.
In the usual construction (see e.g. \cite{Kod,TT,TT2}) one starts with the action $S(x,z)$ of the form (\ref{S-S0}) obeying the Hamilton-Jacobi type equations. The proposition \ref{w-S0-prop} shows that for validness of such a scheme it is sufficient to require the closedness only of the forms (\ref{wjk-S0}) .\par
Finally we note that the formulae (\ref{symm-S0}) and (\ref{Hjk-S0}) imply the existence of the function $F$ such that
(e.g. \cite{KM}) 
\begin{equation}
 \label{Hjk-F-S0}
H^j_k=-\frac{1}{k}\frac{\partial^2 F}{\partial x_j\partial x_k}, \qquad j,k=1,3,5,\dots\ .
\end{equation}
 Consequently the algebraic relations (\ref{Hk111-S0})  become differential equations for $F$ which are equivalent to the well known Hirota equations (see e.g. \cite{Kod,TT,TT2})
\begin{equation}
\label{Hirota-S0} 
\begin{split}
  \frac{\partial^2 F}{\partial x_1\partial x_3}-\frac{3}{2}\left(\frac{\partial F}{\partial x_1^2}\right)^2&=0 , \\
  \frac{\partial^2 F}{\partial x_1\partial x_5}+\frac{5}{2}\left(\frac{\partial F}{\partial x_1^2}\right)^3&=0 , \\
 \dots &
 \end{split}
\end{equation}
 and so on for the BH hierarchy.\par
Solutions of the BH hierarchy or the BH's Hirota equations provide us with the particular class of 2-cocycles (\ref{2-cocy})  given by
\begin{equation}
 \label{2-cocy-S0}
\begin{split}
\psi_0(p_{2j+1},p_{2k+1})=&-\sum_{l=1}  \left( \frac{1}{2(j-l)+1}\frac{\partial^2 \Delta F}{\partial x_{2k+1}\partial x_{2(j-l)+1}} + \frac{1}{2(k-l)+1}\frac{\partial^2 \Delta F}{\partial x_{2j+1}\partial x_{2(k-l)+1}}  
\right) z^{2l}, \\
\psi_0(p_{2j+1},p_{2k+1})=& -\sum_{l=1}  \left( \frac{1}{2(k-l)-1}\frac{\partial^2 \Delta F}{\partial x_{2j+1}\partial x_{2(k-l)-1}} \right) p_{2l+1}, \\
\psi_0(p_{2j+1},p_{2k+1})=& 0, \qquad j,k=0,1,2,3, \dots,
\end{split}
\end{equation}
where $\Delta F$ denotes a variation of the function $F$, for example, $\Delta F =
\sum_{m=1}^\infty \alpha_m \frac{\partial F}{\partial x_{2m+1}}$ where $\alpha_m$ are arbitrary constants. For the 2-coboundary  $f_0(p_{2k+1})$ one has
\begin{equation}
\label{2-cob-S0}
 f_0(p_{2j+1})= \mathcal{D}(z) \Delta F =\sum_{m=1}^\infty \frac{1}{z^{2m+1}} \frac{\partial \Delta F}{\partial x_{2m+1}}
\end{equation}
where $\mathcal{D}(z) $ is the standard vertex operator.
\section{Stratum $\mathbf{\Sigma_2}$: elliptic curve and associated integrable equations}
\label{sec-int-1} 
For the stratum $\Sigma_2$ the system (\ref{ppCp}) is equivalent to the system (\ref{C2g+1}),(\ref{lg}) with $g=1$ and 
$p_{2n}=z^{2n}=\lambda^n$, i.e.
\begin{equation}
\label{C3-S1}
 p_3^2=\lambda^3+u_2\lambda^2+u_1\lambda+u_0,
\end{equation}
and
\begin{equation}
 p_{2m+1}=\left( \lambda^{m-1} +\sum_{k=0}^{m-2} H^{2(m-k)-1}_{-1} \lambda^k\right)p_3
\end{equation}
where
\begin{equation}
\label{ui-C3-S1}
 u_2=2H^3_{-1},\quad u_1=2H^3_{1}+\left(H^3_{-1}\right)^2,\quad u_0=2H^3_{3}+2H^3_{-1}H^3_{1}.
\end{equation}
The associativity condition (\ref{CCuCC}) are reduced to the infinite system of the equations for $H^{2j+1}_k$, the first of which are given by
\begin{equation}
\label{H5j3i-S1}
 \begin{split}
  H^5_{-1}=& H^3_{{1}}-({H^3_{{-1}}})^{2},\\
  H^5_{1}=& -H^3_{{-1}}H^3_{{1}}+H^3_{{3}},\\
  H^5_{3}=& -\frac{1}{2}\,({H^3_{{1}}})^{2}-2\,H^3_{{3}}H^3_{{-1}},\\
 \dots, &
 \end{split}
\end{equation}
and
\begin{equation}
\label{H7j3i-S1}
 \begin{split}
  H^7_{-1}=& -2\,H^3_{{-1}}H^3_{{1}}+H^3_{{3}}+({H^3_{{-1}}})^{3},\\
  H^7_{1}=& -\frac{3}{2}\,({H^3_{{1}}})^{2}+H^3_{{1}}{H^3_{{-1}}}^{2}-2\,H^3_{{3}}H^3_{{-1}},\\
  H^7_{3}=& H^3_{{-1}}({H^3_{{1}}})^{2}+3\,H^3_{{3}}({H^3_{{-1}}})
^{2}-2\,H^3_{{1}}H^3_{{3}},\\
\dots, &
 \end{split}
\end{equation}
and
\begin{equation}
\label{anomaly-S1}
 \begin{split}
  5H^3_5-3H^5_3 =& -(H^3_{{1}})^{2}+H^3_{{3}}H^3_{{-1}},\\
  7H^3_7-3H^7_3 =& \frac{1}{2}\,H^3_{{-1}}({H^3_{{1}}})^{2}-2\,H^3_{{3}}({H^3_{{-1}}})^{2}-H^3_{{1}}H^3_{{3}},\\
  9H^3_9-3H^9_3 =& 3\,H^3_{{3}}({H^3_{{-1}}})^{3}+\frac{3}{2}\,({H^3_{{1}}})^{3},\\
  7H^5_7-5H^7_5 =& H^3_{{3}}({H^3_{{-1}}})^{3}-\frac{3}{2}\,({H^3_{{1}}})^{3}+H^3
_{{1}}H^3_{{3}}H^3_{{-1}}+\frac{1}{2}\,({H^3_{{-1}}})^{2}({H^3_{{1}}})^{2}-({H^3_{{3}}})^{2},\\
\dots & \ .
 \end{split}
\end{equation}
These and other such relations show that there are only three independent elements among $H^{2j+1}_k$. It is convenient  to choose $H^3_{-1}$, $H^3_{1}$, and $H^3_{3}$ as the independent one since these variables first define an elliptic curve (\ref{C3-S1}) and second they provide a simple parameterization of the algebraic variety  $W_{1c}$ defined by the equations (\ref{H5j3i-S1}), (\ref{H7j3i-S1}), and (\ref{anomaly-S1}) and so on. One readily gets
\begin{prop}
 The variety $W_{1c}$ generically is a three dimensional one and the variety $W_1$ is the three parametric family of the coordinate rings for elliptic curves.\par
\end{prop}
In particular the relations (\ref{H5j3i-S1}) define a three dimensional subvariety immersed into the six dimensional euclidean space with coordinates $H^3_{-1}$, $H^3_{1}$, $H^3_{3}$, $H^5_{-1}$, $H^5_{1}$, $H^5_{3}$. The induced metric on this subvariety is 
\begin{equation}
\begin{split}
 ds^2=& \left(1+4\,{y_{{1}}}^{2}+{y_{{2}}}^{2}+4\,{y_{{3}}}^{2}\right)dy_1^2 +
\left(2+{y_{{1}}}^{2}+{y_{{2}}}^{2}\right)dy_2^2+
\left(2+4\,{y_{{1}}}^{2}\right)dy_3^2 \\&
+2\left( -2\,y_{{1}}+y_{{1}}y_{{2}}+2\,y_{{2}}y_{{3}}\right) dy_1dy_2
+2\left( -y_{{2}}+4\,y_{{1}}y_{{3}} \right) dy_1dy_3
+2\left( -y_{{1}}+2\,y_{{1}}y_{{2}} \right) dy_2dy_3
\end{split}
\end{equation}
 where $H^3_{-1}=y_1$, $H^3_{1}=y_2$, $H^3_{3}=y_3$ are chosen as local coordinates . The Riemannian curvature tensor has the following nonzero components 
\begin{equation}
 \begin{split}
  R_{1212}=&\frac{-2-{y_{{2}}}^{2}-16\,y_{{1}}y_{{3}}+4\,y_{{2}}-8\,{y_{{1}}}^{2}-12\,y_
{{1}}y_{{2}}y_{{3}}-8\,{y_{{3}}}^{2}-8\,{y_{{1}}}^{2}y_{{2}}-16\,{y_{{
1}}}^{4}-8\,{y_{{1}}}^{3}y_{{3}}+2\,{y_{{2}}}^{3}}{D}, \\
R_{1213}=&\frac{2\,y_{{1}}y_{{2}}-8\,y_{{1}}+8\,y_{{2}}y_{{3}}+8\,{y_{{1}}}^{2}y_{{3}}
+4\,y_{{1}}{y_{{2}}}^{2}-16\,{y_{{1}}}^{3}+8\,{y_{{1}}}^{3}y_{{2}}}{D},\\
R_{1223}=&\frac{-8-18\,{y_{{1}}}^{2}-8\,{y_{{1}}}^{4}-4\,{y_{{2}}}^{2}-8\,{y_{{1}}}^{2
}y_{{2}}}{D},\\
R_{1313}=&\frac{-16-8\,{y_{{2}}}^{2}-36\,{y_{{1}}}^{2}-16\,{y_{{1}}}^{2}y_{{2}}-16\,{y
_{{1}}}^{4}}{D},
 \end{split}
\end{equation}
where
\begin{equation}
\begin{split} 
D=&4+4\,{y_{{2}}}^{2}+17\,{y_{{1}}}^{2}-4\,y_{{1}}{y_{{2}}}^{2}y_{{3}}+32
\,y_{{1}}y_{{2}}y_{{3}}+16\,{y_{{3}}}^{2}+{y_{{2}}}^{4}+24\,{y_{{1}}}^
{2}{y_{{2}}}^{2}+8\,{y_{{1}}}^{2}y_{{2}}+32\,{y_{{1}}}^{4}y_{{2}}\\
&+16\,
{y_{{1}}}^{6}+4\,{y_{{3}}}^{2}{y_{{1}}}^{2}+24\,{y_{{1}}}^{4}+16\,{y_{
{1}}}^{3}y_{{3}}.
\end{split}
\end{equation}
Tangent $T_{W_{1c}}$ and cotangent $T_{W_{1c}}^*$ bundles are defined by the linear equations
\begin{equation}
\label{T-H5j3i-S1}
 \begin{split}
  \Delta^5_{-1}=& \Delta^3_{{1}}-2{H^3_{{-1}}}\Delta^3_{-1},\\
  \Delta^5_{1}=& -\Delta^3_{{-1}}H^3_{{1}}-H^3_{{-1}}\Delta^3_{{1}}+\Delta^3_{{3}},\\
  \Delta^5_{3}=& -{H^3_{{1}}}{\Delta^3_{{1}}}-2\,\Delta^3_{{3}}H^3_{{-1}}-2\,H^3_{{3}}\Delta^3_{{-1}},\\
 \dots &,
 \end{split}
\end{equation}
and
\begin{equation}
\label{T-H7j3i-S1}
 \begin{split}
  \Delta^7_{-1}=& -2\,\Delta^3_{{-1}}H^3_{{1}}-2\,H^3_{{-1}}\Delta^3_{{1}}+\Delta^3_{{3}}+3({H^3_{{-1}}})^{2}\Delta^3_{{-1}},\\
  \Delta^7_{1}=& -3H^3_1\Delta^3_1+\Delta^3_{{1}}{H^3_{{-1}}}^{2}+2 H^3_{{1}}{H^3_{{-1}}}\Delta^3_{{-1}} 
-2\,\Delta^3_{{3}}H^3_{{-1}}-2\,H^3_{{3}}\Delta^3_{{-1}},\\
  \Delta^7_{3}=& \Delta^3_{{-1}}({H^3_{{1}}})^{2}+2H^3_{{-1}}H^3_1{\Delta^3_{{1}}}
+3\,\Delta^3_{{3}}({H^3_{{-1}}})^{2}+6\,H^3_{{3}}{H^3_{{-1}}}{\Delta^3_{{-1}}}
-2\,\Delta^3_{{1}}H^3_{{3}}-2\,H^3_{{1}}\Delta^3_{{3}},\\
\dots &
 \end{split}
\end{equation}
and so on. \par
The variety $W_{1c}$ is a regular one and hence, the bundles $T_{W_{1c}}$ and $T_{W_{1c}}^*$ are three dimensional one.\par
Now, following the general idea described in the previous section, we will consider a particular subvariety $W_{1c}^I$ for which the three one-forms
\begin{equation}
 \omega_i=H^7_i dx_7+H^5_i dx_5+H^3_i dx_3, \qquad i=-1,1,3,
\end{equation}
 where $x_3$, $x_5$, $x_7$ are some local coordinates in $W_{1c}$, are closed. 
Equivalently $W^I_{1c}$ is a subvariety of $W_{1c}$ such that
\begin{equation}
 d \omega_i \Big{\vert}_{W_{1c}^I} \equiv \left( 
dH^7_i \wedge dx_7+dH^5_i \wedge dx_5+dH^3_i \wedge dx_3\right) \Big{\vert}_{W_{1c}^I}=0.
\end{equation}
The conditions $d\omega_i=0$ locally imply that
\begin{equation}
\label{Hk1-S-S1}
H^3_i=\frac{\partial S_i}{\partial x_3}, \quad
H^5_i=\frac{\partial S_i}{\partial x_5}, \quad
H^7_i=\frac{\partial S_i}{\partial x_7}, \quad i =-1,1,3
\end{equation}
where $S_i$ $(i=-1,1,3)$ are three functions such that $\omega_i=dS_i$. Substitution of these expressions into (\ref{H5j3i-S1}) and (\ref{H7j3i-S1}) give rise to the partial differential equations
\begin{equation}
\label{Eqn-S-H5j3i-S1}
 \begin{split}
  {\frac{\partial S_{-1}}{\partial x_5}}=& \frac{\partial S_1}{\partial x_3}
-\left({\frac{\partial S_{-1}}{\partial x_3}}\right)^{2},\\
  \frac{\partial S_{1}}{\partial x_5}=& -\frac{\partial S_{-1}}{\partial x_3}\frac{\partial S_1}{\partial x_3}+\frac{\partial S_{3}}{\partial x_3},\\
  \frac{\partial S_{3}}{\partial x_5}=& -\frac{1}{2}\,\left({\frac{\partial S_1}{\partial x_3}}\right)^{2}-2\,\frac{\partial S_{3}}{\partial x_3}\frac{\partial S_{-1}}{\partial x_3},\\
 \dots &,
 \end{split}
\end{equation}
and
\begin{equation}
\label{Eqn-S-H7j3i-S1}
 \begin{split}
  {\frac{\partial S_{-1}}{\partial x_7}}=& -2\,\frac{\partial S_{-1}}{\partial x_3}\frac{\partial S_1}{\partial x_3}+\frac{\partial S_{3}}{\partial x_3}+\left({\frac{\partial S_{-1}}{\partial x_3}}\right)^{3},\\
  {\frac{\partial S_{1}}{\partial x_7}}=& -\frac{3}{2}\,\left({\frac{\partial S_1}{\partial x_3}}\right)^{2}+\frac{\partial S_1}{\partial x_3}\left({\frac{\partial S_{-1}}{\partial x_3}}\right)^{2}-2\,\frac{\partial S_{3}}{\partial x_3}\frac{\partial S_{-1}}{\partial x_3},\\
  {\frac{\partial S_{3}}{\partial x_7}}=& 
\frac{\partial S_{-1}}{\partial x_3}\left({\frac{\partial S_1}{\partial x_3}}\right)^{2}
+3\,\frac{\partial S_{3}}{\partial x_3}\left({\frac{\partial S_{-1}}{\partial x_3}}\right)^{2}
-2\,\frac{\partial S_1}{\partial x_3}\frac{\partial S_{3}}{\partial x_3},\\
\dots &,
 \end{split}
\end{equation}
while the comparison with the equations (\ref{T-H5j3i-S1}), (\ref{T-H7j3i-S1}) show that
\begin{equation}
 \Delta^k_i=\frac{\partial \Delta S_i}{\partial x_k}, \qquad i=-1,1,3,\ k=3,5,7,\dots\ .
\end{equation}
\begin{lem}
\label{dwk=0-S1-lem}
 The closedness of the forms (\ref{Hk1-S-S1}) implies that all one-form
\begin{equation}
 \omega_i=H^7_i dx_7+H^5_i dx_5+H^3_i dx_3, \qquad i=-1,1,3
\end{equation}
 for all $\qquad i=-1,1,3,5,7,\dots$ or, equivalently the one-form  
\begin{equation}
 \omega_i=p_7(z) dx_7+p_5(z) dx_5+p_3(z) dx_3,
\end{equation}
are closed.
\end{lem}
Proof is by direct calculation with the use of the algebraic relation (\ref{H5j3i-S1}), (\ref{H7j3i-S1}), (\ref{anomaly-S1}) 
and others. \par 
Thus locally, 
\begin{equation}
\label{wi=dSi-S1}
 \omega(z)=dS(z), \qquad p_k(z)=\frac{\partial S}{\partial x_k},\ k=3,5,7
\end{equation}
where
\begin{equation}
\label{Si-S1}
 S(z)=x_7z^7+x_5z^5+x_3z^3+S_{-1}z+ \sum_{k=0}^\infty\frac{S_{2k+1}}{z^{2k+1}}.
\end{equation}
Moreover one can show that the differential consequences of all other equations defining the subvariety $W_{1c}$ are satisfied due to equations  (\ref{Eqn-S-H5j3i-S1}), (\ref{Eqn-S-H7j3i-S1}) and flows given by these equations commute.\par
So one has 
\begin{prop}
 The subvariety $W_{1c}$ for which one-forms (\ref{Hk1-S-S1}) are closed is characterized (locally) by the compatible system of PDEs (\ref{Eqn-S-H5j3i-S1}), (\ref{Eqn-S-H7j3i-S1}).
\end{prop}
As far the variety $W_1$ is concerned the system (\ref{Eqn-S-H5j3i-S1}), (\ref{Eqn-S-H7j3i-S1}) defines a special family of the coordinate rings for the elliptic curves (\ref{C3-S1}) or, equivalently, a special family of the deformed elliptic curves parameterized by the solutions of the system (\ref{Eqn-S-H5j3i-S1}), (\ref{Eqn-S-H7j3i-S1}). In terms of the coefficients $u_0$, $u_1$, $u_2$ (see (\ref{ui-C3-S1})) defining an elliptic curve, equations (\ref{Eqn-S-H5j3i-S1}), (\ref{Eqn-S-H7j3i-S1}) are given by
\begin{equation}
 \label{Eqn-ui5-S1}
\begin{split}
 \frac{\partial u_2}{\partial x_5}  = & 
- \frac{3}{2}\, \left( {\frac {\partial }{\partial x_{{3}}}}u_{{2}}   \right) u_{{2}}   
+{\frac {\partial }{\partial x_{{3}}}}u_{{1}},   \\
 \frac{\partial u_1}{\partial x_5}  = & {\frac {\partial }{\partial x_{{3}}}}u_{{0}}   
- \frac{1}{2}\, \left( {\frac {\partial }{\partial x_{{3}}}}u_{{1}} \right) u_{{2}}   - 
\left( {\frac {\partial }{\partial x_{{3}}}}u_{{2}} \right) u_{{1}},   \\
 \frac{\partial u_0}{\partial x_5}  = &- \frac{1}{2}\, \left( {\frac {\partial }{\partial x_{{3}}}}u_{{0}} \right) u_{{2}}   
- \left( {\frac {\partial }{\partial x_{{3}}}}u_{{2}} \right) u_{{0}},   
\end{split}
\end{equation}
and
\begin{equation}
 \label{Eqn-ui7-S1}
\begin{split}
 \frac{\partial}{\partial x_7} u_2 = & - \frac{3}{2}\, \left( {\frac {\partial }{\partial x_{{3}}}}u_{{1}} \right) u_{{2}}   
+{\frac {15}{8}}\, \left( {\frac {\partial }{\partial x_{{3}}}}u_{{2}} \right)  \left( u_{{2}}   
 \right) ^{2}- \frac{3}{2}\, \left( {\frac {\partial }{\partial x_{{3}}}}u_{{2}} \right) u_{{1}}   +
{\frac {\partial }{\partial x_{{3}}}}u_{{0}},   \\
 \frac{\partial}{\partial x_7} u_1 = & - \frac{1}{2}\, \left( {\frac {\partial }{\partial x_{{3}}}}u_{{0}}  \right) u_{{2}}   - \frac{3}{2}\,u_{{1}} {\frac {\partial }{\partial x_{{3}}}}u_{{1}}
   + \frac{3}{2}\, \left( {\frac {\partial }{\partial x_{{3}}}}u_{{2}} \right) u_{{1}} 
  u_{{2}}   - \left( {\frac {\partial }{\partial x_{{3}}}}u_{{2}}   \right) u_{{0}}   
+ \frac{3}{8}\, \left( {\frac {\partial }{\partial x_{{3}}}}u_{{1}}   \right)   u_{{2}}^2,\\
 \frac{\partial}{\partial x_7} u_0 = &-u_{{0}}   {\frac {\partial }{\partial x_{{3}}}}u_{{1}}   
- \frac{1}{2}\, \left( {\frac {\partial }{\partial x_{{3}}}}u_{{0}}    \right) u_{{1}}
+ \frac{3}{8}\, \left( {\frac {\partial }{\partial x_{{3}}}}u_{{0}}    \right)  
 u_{{2}}^2
+ \frac{3}{2}\, \left( {\frac {\partial }{\partial x_{{3}}}}u_{{2}}    \right) u_{{0}}   u_{{2}}\  .  
\end{split}
\end{equation}
The system (\ref{Eqn-ui5-S1}) is the well known and well studied dispersionless coupled KdV system (see e.g. \cite{FP,KK,KMAM-10}) while (\ref{Eqn-ui7-S1}) gives its first higher order symmetry. The systems (\ref{Eqn-ui5-S1}) and (\ref{Eqn-ui7-S1})
are integrable hydrodynamical type systems with number of remarkable properties (see e.g. \cite{FP,KK,KMAM-10}): they have infinite set of symmetries and conservation laws, they belong to the infinite hierarchy etc.  Within a different approach \cite{KK} they arose in the Birkhoff stratum $\Sigma_2$ of Gr$^{(2)}$ as the hidden BH equations. We would like to emphasize that in the present context they have a meaning of equations describing a special class of algebraic varieties $W_{1c}^I$. \par
Passing to the infinite dimensional variety $W_1$  one has an infinite dimensional cotangent bundle $T^*_{W_1}$. Hence, one can consider special varieties $W_1^I$ for which the one-forms
\begin{equation}
 \omega_i=\sum_{k=1}^\infty H^{2k+1}_i dx_{2k+1},\qquad i=-1,1,3
\end{equation}
are closed
 or, equivalently, the one-form
\begin{equation}
 \omega(z)=\sum_{k=1}^\infty p_{2k+1}(z)dx_{2k+1}
\end{equation}
where $x_{2k+1}$, $k=1,2,3,\dots$ are local coordinates in $W_1$, is closed. In this case
\begin{equation}
 \omega(z)=dS(z)
\end{equation}
where
\begin{equation}
\label{S-S1}
S(z)=\sum_{m=1}^\infty z^{2m+1} x_{2m+1} +zS_{-1}(x)+\sum_{k=0}^\infty \frac{S_{2k+1}}{z^{2k+1}}
\end{equation}
and $p_j(z)=\frac{\partial S}{\partial x_j}$, $j=3,5,7,\dots$. The action $S(z)$ (\ref{S-S1}) is of the form found by different method in \cite{KK} for the hidden BH hierarchy.
\section{Deformations of moduli for elliptic curves and 2-cocycles}
\label{sec-mod-1} 
The systems (\ref{Eqn-ui5-S1}) and (\ref{Eqn-ui7-S1}) are of particular interest for the theory of deformations of elliptic curves and corresponding Riemann surfaces. Each solution of this system provides us with a nontrivial deformation of the curve (\ref{C3-S1}).\par In terms of the moduli $g_2$ and $g_3$ of an elliptic curve, i.e. in terms of (see e.g. \cite{Sil})
\begin{equation}
\label{g2g3}
 \begin{split}
g_2=& u_1-\frac{1}{3}u_2^2=2H^3_1-\frac{1}{3}(H^3_{-1})^2,\\
g_3=& u_0+\frac{2}{27}u_2^3-\frac{1}{3}u_1u_2=
\frac{2}{3}\,H^3_{{1}}H^3_{{-1}}-{\frac {2}{27}}\,({H^3_{{-1}}})^{3}+2\,H^3_{{3}},
 \end{split}
\end{equation}
equations (\ref{Eqn-S-H5j3i-S1}), (\ref{Eqn-S-H7j3i-S1}) or (\ref{Eqn-ui5-S1}), (\ref{Eqn-ui7-S1}) are of the form (see also \cite{KO-Coiso})
\begin{equation}
\label{g2g3-x5-S1}
\begin{split}
  \frac{\partial g_2}{\partial x_5} =& {\frac {\partial g_{{3}}}{\partial x_3}}
 -\frac{5}{6}\, {\frac {\partial g_{{2}}}{\partial x_3}} u_{{2}}\
 -\frac{2}{3}\,   {\frac {\partial u_{{2}}}{\partial x_3}}   g_{{2}},\\
  \frac{\partial g_3}{\partial x_5} =& -\frac{5}{6}\, {\frac {\partial g_{{3}}}{\partial x_3}}u_{{2}} -\frac{1}{3}
  {\frac{\partial g_{{2}}}{\partial x_3}} g_{{2}} -   {\frac {\partial u_{{2}}}{\partial x_3}}g_{{3}},    \\
  \frac{\partial u_2}{\partial x_5} =& {\frac {\partial g_2}{\partial x_3}}  -\frac{5}{6}\,   {\frac {\partial u_2}{
\partial x_3}}     u_2, 
\end{split}
\end{equation}
and
\begin{equation}
\label{g2g3-x7-S1}
\begin{split}
  \frac{\partial g_2}{\partial x_7} =& -
\frac{7}{6}\,u_{{2}} {\frac {\partial }{\partial x_{
{3}}}}g_{{3}} +{\frac {7}{9}}\,u_{{2}}
  \left( {\frac {\partial }{\partial x_{{3}}}
}u_{{2}}  \right) g_{{2}} 
+{\frac {35}{72}}\, { u_{{2}}} 
  ^{2}{\frac {\partial }{\partial x_{{3}}}}g_{{2}} 
-\frac{3}{2}\,g_{{2}} {\frac {\partial }
{\partial x_{{3}}}}g_{{2}} - 
\left( {\frac {\partial }{\partial x_{{3}}}}u_{{2}} 
 \right) g_{{3}} ,\\
  \frac{\partial g_3}{\partial x_7} =& \frac{7}{6}\,g_{{3}}  
\left( {\frac {\partial }{\partial x_{{3}}}}u_{{2}}  \right) u_{{2}}
 -g_{{3}} 
{\frac {\partial }{\partial x_{{3}}}}g_{{2}} -\frac{5}{6}\,
 \left( {\frac {\partial }{\partial x_{{3}}}}g_{{3}} 
 \right) g_{{2}} +{\frac {35}{72}}\,
 \left( {\frac {\partial }{\partial x_{{3}}}}g_{{3}} 
 \right)  {u_{{2}}} ^{
2}+\frac{2}{9}\, \left( {\frac {\partial }{\partial x_{{3}}}}u_{{2}} 
 \right) {g_{{2}} }^{2}\\&+{\frac {7}{18}}\,u_{{2}} g_{{2}
} {\frac {\partial }{\partial x_{{3}}}}g_{{2}
} ,    \\
  \frac{\partial u_2}{\partial x_7} =& -\frac{7}{6}\,u_{{2}} 
{\frac {\partial }{\partial x_{{3}}}}g_{{2}} 
+{\frac {35}{72}}\, \left( {\frac {\partial }{\partial x_{{3}}}}u_{{2}} 
 \right) { u_{{2}} } ^{2}-\frac{7}{6}\,
 \left( {\frac {\partial }{\partial x_{{3}}}}u_{{2}} 
 \right) g_{{2}} 
+{\frac {\partial }{\partial x_{{3}}}}g_{{3}}  
\end{split}
\end{equation}
or equivalently
\begin{equation}
\label{g2g3Sm1-x5-S1}
\begin{split}
  \frac{\partial g_2}{\partial x_5} =& {\frac {\partial g_{{3}}}{\partial x_3}}
 -\frac{5}{3}\, {\frac {\partial g_{{2}}}{\partial x_3}}  {\frac {\partial S_{{-1}}}{\partial x_3}} \
 -\frac{4}{3}\,   {\frac {\partial^2 S_{{-1}}}{\partial x_3^2}}   g_{{2}},\\
  \frac{\partial g_3}{\partial x_5} =& -\frac{5}{3}\, {\frac {\partial g_{{3}}}{\partial x_3}}{\frac {\partial S_{{-1}}}{\partial x_3}} 
-\frac{1}{3}  {\frac{\partial g_{{2}}}{\partial x_3}} g_{{2}} -  2 {\frac {\partial^2 S_{{-1}}}{\partial x_3^2}}g_{{3}},    \\
  \frac{\partial S_{-1}}{\partial x_5} =& \frac{1}{2} g_2  -\frac{5}{6}\,   \left({\frac {\partial S_{-1}}{
\partial x_3}}     \right)^2,
\end{split}
\end{equation}
and
\begin{equation}
\label{g2g3Sm1-x7-S1}
\begin{split}
  \frac{\partial g_2}{\partial x_7} =& -
\frac{7}{3}\,{\frac {\partial S_{-1}}{
\partial x_3}} {\frac {\partial }{\partial x_{
{3}}}}g_{{3}} +{\frac {28}{9}}\,{\frac {\partial S_{-1}}{\partial x_3}}
  \left( {\frac {\partial^2 }{\partial x_{{3}}^2}}S_{{-1}}  \right) g_{{2}} 
+{\frac {35}{18}}\, \left( {\frac {\partial S_{-1}}{
\partial x_3}} \right) ^{2}{\frac {\partial }{\partial x_{{3}}}}g_{{2}} 
-3\,g_{{2}} {\frac {\partial }
{\partial x_{{3}}}}g_{{2}} \\&- 2\left( {\frac {\partial^2 }{\partial x_{{3}}^2}} S_{{-1}} \right) g_{{3}} ,\\  
  \frac{\partial g_3}{\partial x_7} =& 
\frac{7}{3}\,g_{{3}}  \left( {\frac {\partial }{\partial x_{{3}}}}u_{{2}}  \right) {\frac {\partial S_{-1}}{
\partial x_3}}
 -g_{{3}} {\frac {\partial }{\partial x_{{3}}}}g_{{2}} 
-\frac{5}{6}\, \left( {\frac {\partial }{\partial x_{{3}}}}g_{{3}} \right) g_{{2}} 
+{\frac {35}{18}}\, \left( {\frac {\partial }{\partial x_{{3}}}}g_{{3}} 
 \right)  \left( {\frac {\partial S_{-1}}{\partial x_3}}  \right) ^{2}
\\&+\frac{4}{9}\, \left( {\frac {\partial^2 }{\partial x_{{3}}^2}}S_{{-1}} \right)  {g_{{2}}} ^{2}
+{\frac {7}{9}}\, g_{{2} } {\frac {\partial S_{-1}}{\partial x_3}}{\frac {\partial }{\partial x_{{3}}}}g_{{2}
} ,    \\
  \frac{\partial S_{-1}}{\partial x_7} =& 
-\frac{7}{6}\,g_{{2}} {\frac {\partial }{\partial x_{{3}}}}S_{{-1}} 
+{\frac {35}{54}}\, \left( {\frac {\partial }{\partial x_{{3}}}}S_{{-1}} 
 \right)^3  
+\frac{1}{2}g_{{3}} . 
\end{split}
\end{equation}
Deformations of the discriminant  $\Delta=-16\left( 2g_2^3+27g_3^2\right)$ of an elliptic curve (\ref{C3-S1}) are defined by the equation
\begin{equation}
\label{DDelta-S1}
\begin{split}
 \frac{\partial \Delta}{\partial x_5}=&
-192\, \left( g_{{2}}   \right) ^{2}{\frac {
\partial }{\partial x_{{3}}}}g_{{3}}  +128\,
 \left( g_{{2}}   \right) ^{3}{\frac {
\partial }{\partial x_{{3}}}}u_{{2}}  +160\,
 \left( g_{{2}}   \right) ^{2} u_{{2}} 
 {\frac {\partial }{\partial x_{{3}}}}g_{{2}} 
 +720\,g_{{3}}  u_{{2}}
  {\frac {\partial }{\partial x_{{3}}}}g_{{3}}
  \\& +864\, \left( {\frac {\partial }{\partial x_
{{3}}}}u_{{2}}   \right)  \left( g_{{3}}
   \right) ^{2}+288\,g_{{3}} 
 g_{{2}}  {\frac {\partial }{\partial 
x_{{3}}}}g_{{2}}\ .  
\end{split}
\end{equation}
Deformations of the moduli $g_2$, $g_3$ and the elliptic curve described by the equations  (\ref{g2g3Sm1-x5-S1})-(\ref{DDelta-S1}) exhibit a rich structure. In this deformations the process $\Delta \to 0$ or 
$\Delta=0\ \to \Delta \neq 0$ may occur. Such deformations describe the degeneration and desingularization of an elliptic curve. Deformations of such type associated with some particular solutions of equations (\ref{g2g3Sm1-x5-S1})-(\ref{DDelta-S1}) have been studied in \cite{KO-Ham}.\par
Equations (\ref{Eqn-S-H5j3i-S1}), (\ref{Eqn-S-H7j3i-S1}) or (\ref{Eqn-ui5-S1}), (\ref{Eqn-ui7-S1}) provide us with a special class of 2-cocycles and 2-coboundaries related to an elliptic curve, namely 
\begin{equation}
\label{2cocy-S1}
\begin{split}
 \psi_1(p_{2n},p_{2m})=&0, \\
 \psi_1(p_{2n+1},p_{2m+1})=& 
\sum_{k=-1}^{m}\frac{\partial \Delta S_{2k+1}}{\partial x_{2n+1}} p_{2(m-k)} 
+\sum_{k=-1}^{n} \frac{\partial \Delta S_{2k+1}}{\partial x_{2m+1}}  p_{2(n-k)}\\
&+\left(\frac{\partial \Delta S_{-1}}{\partial x_{2n+1}} H^{2m+1}_{-1}
+\frac{\partial \Delta S_{-1}}{\partial x_{2m+1}} H^{2n+1}_{-1}\right)p_2 \\&
+\frac{\partial \Delta S_{-1}}{\partial x_{2m+1}} H^{2n+1}_{1}
+\frac{\partial \Delta S_{1}}{\partial x_{2m+1}} H^{2n+1}_{-1}
+\frac{\partial \Delta S_{-1}}{\partial x_{2n+1}} H^{2m+1}_{1}
+\frac{\partial \Delta S_{1}}{\partial x_{2n+1}} H^{2m+1}_{-1}, \\
 \psi_1(p_{2n},p_{2m+1})=&\sum_{k=-1}^{n-2} \frac{\partial \Delta S_{2k+1}}{\partial x_{2m+1}} p_{2(n-k)-1}, 
\end{split}
\end{equation}
and
\begin{equation}
\label{2cob-S1}
 f_1(p_{2j+1})=\frac{\partial \Delta S_{-1}}{\partial x_{2j+1}} 
+\sum_{m=0}^\infty \frac{1}{z^{2m+1}} \frac{\partial \Delta S_{2m+1}}{\partial x_{2j+1}},
\end{equation}
where $\Delta S_{2j+1}$ denote variations of $S_{2j+1}$. As a special case one has $\Delta S_{2j+1}=\sum_{m=1}^\infty \frac{\partial \Delta S_{2j+1}}{\partial x_{2m+1}}$ where $\alpha_m$ are arbitrary constants.\par
These formulae are quite similar to those for the big  cell written in terms of the corresponding $S_{2k+1}$. In the big cell, due to the relations (\ref{symm-S0}) one can go further and express all $S_{2k+1}$ as the derivatives of a single function $F$ and get Hirota equations (\ref{Hirota-S0}). Such a property is not valid, in general, for the variety $W_1$ in the stratum $\Sigma_2$. Indeed one has relations (\ref{anomaly-S1}) instead of (\ref{symm-S0}). So, even all
$H^{2j+1}_{2k+1}=\frac{\partial S_{2k+1}}{\partial x_{2j+1}}$ the relations (\ref{anomaly-S1}), in contrast to the relations  (\ref{symm-S0}) apparently, do not imply the existence of a single function $F$ such that
$H^{2j+1}_{2k+1}=\frac{\partial^2 S}{\partial x_{2j+1} \partial x_{2k+1}}$. This fact supports the observation made in the paper \cite{KK}.  
\section{Deformations of elliptic curves with a fixed point and dispersionless NLS equation}
\label{sec-NLS-1} 
Particular subvarieties in $W_1$ and $W_{1c}$ and corresponding reductions of the system (\ref{Eqn-ui5-S1}), (\ref{Eqn-ui7-S1}) are of interest too. The simplest corresponds to the constraint $u_0=0$ or
\begin{equation}
 \label{constr-NLS-S1}
H^3_3+H^3_{-1}H^3_{1}=0.
\end{equation}
The elliptic curve (\ref{C3-S1}) in this case assumes the form 
\begin{equation}
 \label{C3-NLS-S1}
p_3^2=\lambda^3+u_2\lambda^2+u_1\lambda
\end{equation}
which corresponds to vanishing of one of roots for the cubic polynomial in the r.h.s. of (\ref{C3-S1}). Under the constraint (\ref{constr-NLS-S1}) the subvariety $W_{1c}$ becomes two-dimensional and $H^3_{-1}$, $H^3_1$ can be chosen as the local coordinates on it. Consequently a natural analog of the closedness condition discussed in the previous section is given by
\begin{equation}
\label{wi-NLS-S1}
 d\omega_i=0, \qquad \omega_i=H^5_idx_5+H^3_idx_3, \qquad i=-1,1,3
\end{equation}
under the constraint (\ref{constr-NLS-S1}).
Similar to the Lemma \ref{dwk=0-S1-lem} one can show that the condition (\ref{wi-NLS-S1}) implies the closedness of the form 
\begin{equation}
 \omega(z)=p_5(z)dx_5+p_3(z)dx_3
\end{equation}
and, then, validness of the formulae (\ref{wi=dSi-S1}), (\ref{Si-S1}) with $x_7=0$.\par
The condition (\ref{constr-NLS-S1}) gives rise to the equations
\begin{equation}
 \label{constr-Eqn-ui-S1}
\begin{split}
 \frac{\partial u_2}{\partial x_5}  = & 
- \frac{3}{2}\, \left( {\frac {\partial }{\partial x_{{3}}}}u_{{2}}   \right) u_{{2}}   
+{\frac {\partial }{\partial x_{{3}}}}u_{{1}},   \\
 \frac{\partial u_1}{\partial x_5}  = &    
- \frac{1}{2}\, \left( {\frac {\partial }{\partial x_{{3}}}}u_{{1}} \right) u_{{2}}   - 
\left( {\frac {\partial }{\partial x_{{3}}}}u_{{2}} \right) u_{{1}}.  
\end{split}
\end{equation}
This system describes deformations of the elliptic curve (\ref{C3-NLS-S1}) for which the origin ($p_3=x=0$) is the fixed point.\par
At $u_0=0$ the definition (\ref{g2g3}) implies that
\begin{equation}
 \label{u2-cub-NLS}
\left(\frac{u_2}{3} \right)^3+ \frac{u_2}{3}g_2 +g_3=0.
\end{equation}
Hence, equations (\ref{constr-Eqn-ui-S1}) or (\ref{g2g3-x5-S1}) for moduli $g_2$ and $g_3$ become
\begin{equation}
\label{g2g3-NLS-S1}
\begin{split}
  \frac{\partial g_2}{\partial x_5} =& {\frac {\partial g_{{3}}}{\partial x_3}}
 -\frac{5}{6}\, {\frac {\partial g_{{2}}}{\partial x_3}} u_{{2}}\
 -\frac{2}{3}\,   {\frac {\partial u_{{2}}}{\partial x_3}}   g_{{2}},\\
  \frac{\partial g_3}{\partial x_5} =& -\frac{5}{6}\, {\frac {\partial g_{{3}}}{\partial x_3}}u_{{2}} -\frac{1}{3}
  {\frac{\partial g_{{2}}}{\partial x_3}} g_{{2}} -   {\frac {\partial u_{{2}}}{\partial x_3}}g_{{3}}, 
\end{split}
\end{equation}
where $u_2$ is a root of the cubic equation (\ref{u2-cub-NLS}). The original system (\ref{g2g3-x5-S1}) is compatible with the constraint (\ref{u2-cub-NLS}) which is equivalent to $u_0=0$. Another form of the system (\ref{g2g3-x5-S1}) 
under the constraint (\ref{u2-cub-NLS})is given by the system
\begin{equation}
\label{g2u2-NLS-x5-S1}
\begin{split}
  \frac{\partial g_2}{\partial x_5} =& -\frac{1}{9}u_2^2{\frac {\partial u_{{2}}}{\partial x_3}}
 -\frac{7}{6}\, {\frac {\partial g_{{2}}}{\partial x_3}} u_{{2}}\
 -{\frac {\partial u_{{2}}}{\partial x_3}}   g_{{2}},\\
  \frac{\partial u_2}{\partial x_5} =& {\frac {\partial g_2}{\partial x_3}}  -\frac{5}{6}\,   {\frac {\partial u_2}{
\partial x_3}}     u_2 .
\end{split}
\end{equation}
Solving this system, one reconstructs $g_3=-\left(\frac{u_2}{3} \right)^3- \frac{u_2}{3}g_2$.\par
For the discriminant $\Delta=16u_1^2(u_2^2-4u_1)$ one has
\begin{equation}
\label{DDelta-NLS-S1}
\begin{split}
 \frac{\partial \Delta}{\partial x_5}=&
{\frac {592}{3}}\, \left( u_2 g_{{2}}   \right) ^{2} \left( {\frac {\partial }{\partial x_{{3}}}} u_{{2}} 
  \right)  +192\, \left( g_{{2}}   \right) ^{3}{\frac 
{\partial }{\partial x_{{3}}}}u_{{2}}  +128\,
 \left( g_{{2}}   \right) ^{2}u_{{2}} 
 {\frac {\partial }{\partial x_{{3}}}}g_{{2}} 
 +{\frac {112}{27}}\, \left( u_{{2}} 
  \right) ^{6}{\frac {\partial }{\partial x_{{3}}}}u_{{2}} \\&
  +{\frac {512}{9}}\, \left( u_{{2}}   \right) ^{4} 
\left( {\frac {\partial }{\partial x_{{3}}}}u_{{2}}   \right) g_{{2}} 
 +{\frac {80}{9}}\, \left( u_{{2}}  
 \right) ^{5}{\frac {\partial }{\partial x_{{3}}}}g_{{2}} 
 +{\frac {208}{3}}\, \left( u_{{2}} 
  \right) ^{3}g_{{2}}  {\frac {
\partial }{\partial x_{{3}}}}g_{{2}} .
\end{split}
\end{equation}
We emphasize that the elliptic curve (\ref{C3-NLS-S1}) generically is not singular and it remains almost everywhere regular under deformations given by equations (\ref{constr-Eqn-ui-S1})-(\ref{DDelta-NLS-S1}). There are two obvious constraints, namely $u_1=0$ and $u_2^2=4u_1$  under which the curve (\ref{C3-NLS-S1}) becomes singular ($\Delta=0$). Under both these constraints the system  (\ref{constr-Eqn-ui-S1}) is reduced to the BH equation. So the BH equation describes deformations of the degenerate plane cubic in agreement with the observation made in \cite{KO}. Thus the systems (\ref{constr-Eqn-ui-S1}) or (\ref{g2g3-NLS-S1}) are of importance for the theory of elliptic curves.\par
In fact the system (\ref{constr-Eqn-ui-S1}) is a well-known one in the theory of dispersionless integrable systems. It is the so called dispersionless Jaulent-Miodek system (see e.g. \cite{KMAM-10}). Under the change of the dependent variables
\begin{equation}
 u=-u_2, \qquad v=-u_1+\frac{1}{4}u_2^2
\end{equation}
and $x_3=x$, $x_5=-t$ it becomes the 1-layer Benney system
\begin{equation}
\label{NLS-S1}
\begin{split}
 \frac{\partial u}{\partial t} +u \frac{\partial u}{\partial x} + \frac{\partial v}{\partial x} =& 0, \\
 \frac{\partial v}{\partial t} + \frac{\partial }{\partial x}(uv)=&0,
\end{split}
\end{equation}
 which describes long waves on the shallow water \cite{Ben}. Moreover, the system (\ref{NLS-S1}) is the quasiclassical limit \cite{Zak} of the famous nonlinear Schroedinger (NLS) equation
\begin{equation}
\label{NLS}
 i\epsilon \frac{\partial}{\partial t}\psi +\frac{\epsilon^2}{2}\frac{\partial^2}{\partial x^2}\psi +|\psi|^2\psi=0
\end{equation}
as
\begin{equation}
 \psi=Ae^{\frac{i}{\epsilon}S}, \qquad u=S_x,\ v=-A^2
\end{equation}
and $\epsilon \to 0$.\par
The NLS equation (\ref{NLS}) and its quasiclassical limit arise in the numerous nonlinear phenomena in physics and problems in mathematics (see e.g. \cite{ZMNP,AS}). However its relevance to the deformation theory for elliptic curves seems has not been mentioned before.\par
In addition to a number of remarkable properties typical for integrable hydrodynamical type equations the system (\ref{NLS-S1}) implies the existence of a single function $\phi$ such that it is equivalent to the single Hirota type equation. To demonstrate this it is convenient to use the system (\ref{Eqn-S-H5j3i-S1}) under the constraint (\ref{constr-NLS-S1}), i.e.
\begin{equation}
 \frac{\partial S_3 }{\partial x_3 }+\frac{\partial S_{-1} }{\partial x_3 }\frac{\partial S_1 }{\partial x_3 }=0
\end{equation}
that is
\begin{equation}
\label{Si-NLS-S1}
 \begin{split}
  \frac{\partial S_{-1} }{\partial x_5 }=&\frac{\partial S_1 }{\partial x_3 }-\left( \frac{\partial S_{-1} }{\partial x_3 } \right)^2,\\
\frac{\partial S_1 }{\partial x_5 }=&-2\frac{\partial S_1 }{\partial x_3 }\frac{\partial S_{-1} }{\partial x_3 }.
 \end{split}
\end{equation}
Differentiating the first equation (\ref{Si-NLS-S1}) with respect to $x_3$ and expressing $\frac{\partial S_{-1} }{\partial x_3 }$ in terms of $S_1$, using the second equation of (\ref{Si-NLS-S1}), one gets
\begin{equation}
\label{S1-NLS-S1}
 \frac{\partial^2 S_1}{\partial x_5^2}=\frac{\partial}{\partial x_3}\left( -\left(\frac{\partial S_1}{\partial x_3}\right)^2+
\frac{\left(\frac{\partial S_1}{\partial x_5}\right)^2}{\frac{\partial S_1}{\partial x_3}}
 \right).
\end{equation}
This equation implies the existence of the function $\phi$ such that 
\begin{equation}
\label{S1-phi-S1}
 S_1=\frac{\partial \phi}{\partial x_3}.
\end{equation}
In terms of $\phi$ equation (\ref{S1-NLS-S1}) (choosing vanishing integration constants) is
\begin{equation}
\label{phi-NLS-S1}
 \phi_{x_3x_3}\phi_{x_5x_5} - (\phi_{x_3x_5})^2 + (\phi_{x_3x_3})^3 =0.
\end{equation}
where $\phi_{x_jx_k}=\frac{\partial^2 \phi}{\partial x_j\partial x_k}$.\par
Thus one can construct solutions of the system (\ref{Si-NLS-S1}) solving Hirota (and Hessian) type equation (\ref{phi-NLS-S1}) and using the formula (\ref{S1-phi-S1}) and $\frac{\partial S_{-1}}{\partial x_3}=-\frac{1}{2}\frac{\phi_{x_3x_5}}{\phi_{x_3x_3}}$. Solutions of the system (\ref{constr-Eqn-ui-S1}) are given by
\begin{equation}
 u_2=-\frac{\phi_{x_3x_5}}{\phi_{x_3x_3}}, 
\qquad u_1=2 \phi_{x_3x_3}+\frac{1}{4}\frac{\phi_{x_3x_5}^2}{\phi_{x_3x_3}^2}
\end{equation}
and for moduli $g_2$, $g_3$ one has
\begin{equation}
 \begin{split}
g_2=& 2\phi_{x_3x_3}-\frac{1}{12}\frac{\phi_{x_3x_5}^2}{\phi_{x_3x_3}^2},\\
g_3=& \frac{2}{3} \phi_{x_3x_5}-\frac{1}{108}\frac{\phi_{x_3x_5}^3}{\phi_{x_3x_3}^3},
 \end{split}
\end{equation}
while the solution of the 1-layer Benney system (\ref{NLS-S1})  are ($t=-x_5$) 
\begin{equation} 
 u= \frac{\phi_{x_3x_5}}{\phi_{x_3x_3}}, \qquad v= -2\phi_{x_3x_3}.
\end{equation}
One can get the equation (\ref{phi-NLS-S1}) directly from the 1-layer Benney system (\ref{NLS-S1}) too. In fact, the second equation (\ref{NLS-S1})
 implies the existence of the function $\tilde{\phi}$ such that $v=2\tilde{\phi}_x$ and, hence, $u=-\frac{\tilde{\phi}_t}{\tilde{\phi}_x}$. Substituting these expressions into the first equations (\ref{NLS-S1}), 
one gets
\begin{equation}
\label{tilde-phi-NLS-S1}
 \tilde{\phi}_{tt}=\left( \frac{\tilde{\phi}_t^2}{\tilde{\phi}_x}+\tilde{\phi}_{x}^2\right)_x.
\end{equation}
This implies that $\tilde{\phi}=\overline{\phi}_x$ and equation (\ref{tilde-phi-NLS-S1}) becomes
\begin{equation}
 \overline{\phi}_{tt}=\frac{\overline{\phi}_{xt}^2}{\overline{\phi}_{xx}}+\overline{\phi}_{xx}^2
\end{equation}
that coincides with (\ref{phi-NLS-S1}) modulo substitution $x=-x_3$, $\phi=-\overline{\phi}$.\par
Thus, the Hirota equation (\ref{phi-NLS-S1}) governs deformations of the elliptic curve (\ref{C3-NLS-S1}). It should be relevant also to the study of the quasiclassical limit of the NLS equation. Equation (\ref{phi-NLS-S1}) has several interesting properties. For example, it is invariant under the scale transformations 
\begin{equation}
 \label{scale-inv}
x_3 \to \rho x_3, \qquad x_5 \to \rho^2 x_5, \qquad \phi \to \phi.  
\end{equation}
Hence, it admits self-similar solutions 
\begin{equation}
 \phi=f\left( \frac{x_3^2}{x_5}\right)
\end{equation}
for which it is reduced to the following ODE
\begin{equation}
 \label{ODE-ss-NLS}
y^2 \varphi \varphi' +4 (\varphi +2y\varphi')^3=0
\end{equation}
where $y=\frac{x_3^2}{x_5}$ and $\varphi=\frac{\partial f}{\partial y}$. 
One can show that the only monomial solution of the equation (\ref{ODE-ss-NLS}) is $\varphi=-\frac{1}{108}y$ . This implies that $H^3_1=-\frac{1}{18}\frac{x_3^2}{x_5^2}$ and $H^3_{-1}=\frac{1}{3}\frac{x_3}{x_5}$ and the corresponding family of elliptic curves is degenerate and it is given by
\begin{equation}
 p_3^2=\lambda^3+\frac{2 x_3}{3 x_5}\lambda^2.
\end{equation}
Solutions of equations (\ref{phi-NLS-S1}) provide us with a particular class of 2-cocycles $\psi_1^{dNLS}$ and 2-coboundaries $f_1^{dNLS}$ defined by the formulae (\ref{2cocy-S1}), (\ref{2cob-S1}) under the reduction (\ref{constr-NLS-S1}), for example, 
\begin{equation}
 \begin{split}
  \psi_1^{dNLS}(p_3,p_3)=&(-\Delta u)  \lambda^2 +\left(-\Delta v +\frac{1}{2}  u\Delta u\right) \lambda \\
=& \left(-\frac{(\Delta \phi)_{x_3x_5}}{\phi_{x_3x_3}}+  \frac{\phi_{x_3x_5}}{\phi_{x_3x_3}^2}(\Delta \phi)_{x_3x_3}\right) \lambda^2 \\&
+ \left( 
2(\Delta \phi)_{x_3x_3}
\frac{1}{2}\frac{\phi_{x_3x_5}}{\phi_{x_3x_3}^2}(\Delta \phi)_{x_3x_5}
-\frac{1}{2}\frac{\phi_{x_3x_5}^2}{\phi_{x_3x_3}^3}(\Delta \phi)_{x_3x_3}
\right)
\lambda.
 \end{split}
\end{equation}
One can refer to such 2-cocycles as dNLS 2-cocycles.
\section{Hyperelliptic curves in $\mathbf{W_g}$ and dispersionless coupled KdV equations}
\label{sec-int-g} 
For the strata $\Sigma_{2g}$ ($g>1$) the variety $W_g$ is defined by the relation (\ref{C2g+1}) and (\ref{lg}) and associativity conditions \cite{KO-p1} 
\begin{equation}
\label{H2n-S2g+1}
\begin{split}
 H^{2m}_k =& 0, \qquad m =0,1,2, \dots, \ k= -2g+2, -2g+4, \dots, -2, 0, 1, 2, 3, \dots, \\
H^{2m+1}_{2k} =& 0, \qquad m = 0,1,2, \dots \ k = -g , -g+1, -g+2, \dots 
\end{split}
\end{equation}
and
\begin{equation}
\label{H2n+1-S2g+1}
\begin{split}
&H^{2j+1}_{2(l+k)+1}-H^{2(j+k)+1}_{2l+1}-\sum_{s=-g}^{k-1}H^{2j+1}_{2s+1}H^{2(k-s)-1}_{2l+1}=0, \\
&H^{2j+1}_{2(l+k)+1}+H^{2k+1}_{2(l+j)+1}
+\sum_{s=-g}^{-1} H^{2j+1}_{2s+1}H^{2k+1}_{2(l-s)-1}
+\sum_{r=-g}^{-1} H^{2k+1}_{2r+1}H^{2j+1}_{2(l-r)-1}
+\sum_{s=0}^{l-g} H^{2j+1}_{2s+1}H^{2k+1}_{2(l-s)-1}=0.
\end{split}
\end{equation}
and
\begin{equation}
p_{2g+1}^2 = \lambda^{2g+1} +\sum_{k=0}^{2g} u_k \lambda^k
\end{equation}
where the coefficients $u_k$ in (\ref{C2g+1}) can be obtained from
\begin{equation}
 \label{uC-S2n}
\begin{split}
 p_{2g+1}^2 = \lambda^{2g+1} +2\sum_{s=0}^{2g} H^{2g+1}_{2(g-s)+1} \lambda^s
+\sum_{k=-g}^{g+1} \sum_{s=0}^{g-k-1} H^{2g+1}_{2k+1}H^{2g+1}_{-2(s+k)-1} \lambda^s.
\end{split}
\end{equation}
\begin{lem}
 The subvariety $W_{gc}$ of the coefficients $H^j_k$ has dimension $2g+1$.
\end{lem}
{\bf Proof} We first observe that a hyperelliptic curve (\ref{C2g+1}) is parameterized by $2g+1$ variables $H^{2g+1}_{-(2g-1)}$, $H^{2g+1}_{-(2g-3)}$, $\dots$, $H^{2g+1}_{1}$, $\dots$, $H^{2g+1}_{2g+1}$. Then, evaluating coefficients in front of $\frac{1}{z^{2k+1}}$, $k=0,1,2,\dots$ in equation (\ref{C2g+1}), one concludes that all of them are certain polynomials of these $2g+1$ variables. 
 For instance,
 \begin{equation}
  H^{2g+1}_{2g+3}=-\frac{1}{2}(H^{2g+1}_{1})^2-\sum_{j=1}^g H^{2g+1}_{2j+1} H^{2g+1}_{1-2j} 
\end{equation}
Further, the part of the conditions  (\ref{H2n-S2g+1}), (\ref{H2n+1-S2g+1}) encoded in the relations (\ref{lg}) or, equivalently in the relations 
\begin{equation}
\label{p2g+3}
 p_{2(g+k)+3}-z^2p_{2(g+k)+1}+H^{2(g+k)+1}_{-(2g-1)}p_{2g+1}=0, \qquad k=g,g+1,g+2,\dots
\end{equation}
allows us to express recursively all the variables $H^{2k+3}_j$, $k=g,g+1,g+2,\dots$ in terms of $H^{2g+1}_j$ and, hence, in terms of  $H^{2g+1}_j$ with $j=-(2g-1),-(2g-3),\dots,1,\dots,2g+1$. In particular one has
\begin{equation}  
\label{Halgfin-S2n}
\begin{split}
& H^{2g+3}_{-(2n-1)}- H^{2g+1}_{-(2n-3)} + H^{2g+1}_{2g-1} H^{2g+1}_{2n-1}=0, \qquad -g+1 \leq n \leq g, \\
& H^{2g+3}_{2g+1}- H^{2g+1}_{2g+3}(\{H^{2g+1}_{-2i+1}\}_{i=-g \dots g}) + H^{2g+1}_{2g-1} H^{2g+1}_{2g+1}=0,
\end{split}
\end{equation}
and 
\begin{equation}
\label{H-P-S2n}
 H^{2(g+k)+1}_{-(2n-1)}=P^k_n(\{H^{2g+1}\})
\end{equation}
where $P^k_n$ are certain polynomials of $2g+1$ variables $\{H^{2g+1}\}=\{H^{2g+1}_{1-2g},H^{2g+1}_{3-2g},H^{2g+1}_{5-2g} \dots,H^{2g+1}_{1+2g}\}$. $\square$ \par
For example at $g=2$ one has the set $\{H^5\}=\{H^5_{-3},H^5_{-1},H^5_{1},H^5_{3},H^5_{5}\}$ and the first of the relations
(\ref{Halgfin-S2n}), (\ref{H-P-S2n}) are
\begin{equation}
\label{H7i-S4} 
\begin{split}
H^7_{-3}=& H^5_{{-1}}-{H^5_{{-3}}}^{2},\\
H^7_{-1}=& H^5_{{1}}-H^5_{{-3}}H^5_{{-1}},\\
H^7_{1}=&  H^5_{{3}}-H^5_{{-3}}H^5_{{1}},\\
H^7_{3}=&  H^5_{{5}}-H^5_{{3}}H^5_{{-3}},\\
H^7_{5}=&  -H^5_{{-1}}H^5_{{3}}-\frac{1}{2}\,{H^5_{{1}}}^{2}-2\,H^5_{{5}}H^5_{{-3}},
 \end{split}
\end{equation}
and
\begin{equation}
\label{H9i-S4} 
\begin{split}
H^9_{-3}=& H^5_{{1}}-2\,H^5_{{-3}}H^5_{{-1}}+{H^5_{{-3}}}^{3},\\
H^9_{-1}=& -{H^5_{{-1}}}^{2}+H^5_{{-1}}{H^5_{{-3}}}^{2}
+H^5_{{3}}-H^5_{{-3}}H^5_{{1}},\\
H^9_{1}=&  H^5_{{5}}-H^5_{{3}}H^5_{{-3}}-H^5_{{1}}H^5_{{-1}}
+H^5_{{1}}{H^5_{{-3}}}^{2},\\
H^9_{3}=&  -2\,H^5_{{-1}}H^5_{{3}}+H^5_{{3}}{H^5_{{-3}}}^{2}-
\frac{1}{2}\,{H^5_{{1}}}^{2}-2\,H^5_{{5}}H^5_{{-3}},\\
H^9_{5}=& -2\,H^5_{{-1}}H^5_{{5}}
+3\,H^5_{{5}}{H^5_{{-3}}}^{2}+2\,H^5_{{-3}}H^5_{{-1}}H^5_{{3}}
+H^5_{{-3}}{H^5_{{1}}}^{2}-H^5_{{3}}H^5_{{1}}.
 \end{split}
\end{equation}
The relations (\ref{H7i-S4}), (\ref{H9i-S4}) define the five dimensional algebraic varieties as the intersection of a very special quadrics.\par
So, one has
\begin{prop}
 Variety $W_g$ represents a $2g+1$ dimensional family of the coordinate rings of the deformed hyperelliptic curves (\ref{C2g+1}) of genus $g$. 
\end{prop}
 Tangent and cotangent spaces of $W_{gc}$ are also $2g+1$-dimensional.  They are defined by the linearized versions of the relations (\ref{H2n-S2g+1}), (\ref{H2n+1-S2g+1}) or (\ref{Halgfin-S2n}), (\ref{H-P-S2n}). For example, the part corresponding to the relations (\ref{H7i-S4}) at $g=2$ is given by
\begin{equation}
\label{D-H7i-S4} 
\begin{split}
\Delta^7_{-3}=& \Delta^5_{{-1}}-2{H^5_{{-3}}}{\Delta^5_{{-3}}},\\
\Delta^7_{-1}=& \Delta^5_{{1}}-\Delta^5_{{-3}}H^5_{{-1}}-H^5_{{-3}}\Delta^5_{{-1}},\\
\Delta^7_{1}=&  \Delta^5_{{3}}-\Delta^5_{{-3}}H^5_{{1}}-H^5_{{-3}}\Delta^5_{{1}},\\
\Delta^7_{3}=&  \Delta^5_{{5}}-\Delta^5_{{3}}H^5_{{-3}}-H^5_{{3}}\Delta^5_{{-3}},\\
\Delta^7_{5}=&  -\Delta^5_{{-1}}H^5_{{3}}-H^5_{{-1}}\Delta^5_{{3}} 
-H^5_1{\Delta^5_{{1}}} 
-2\,\Delta^5_{{5}}H^5_{{-3}}-2\,H^5_{{5}}\Delta^5_{{-3}}.
 \end{split}
\end{equation}
\begin{defi}
 Variety $W_{gc}^I$ is the subvariety of $W_{gc}$ for which 
 $2g+1$ one-forms
\begin{equation}
\label{wi-S2n}
 \omega_i=H^{2g+1}_idx_{2g+1}+H^{2g+3}_idx_{2g+3}+\dots+H^{6g+1}_idx_{6g+1},\qquad i=1-2g,3-2g,\dots,1,\dots,2g+1,
\end{equation}
 where $x_{2g+1},x_{2g+3},\dots,x_{6g+1}$ are local coordinates in $W_{gc}$, are closed.
\end{defi}
\begin{lem}
 The closedness of the forms (\ref{wi-S2n}) imply the closedness of all form of type (\ref{wi-S2n}) with all $i=2g+3,2g+5,\dots$ .
\end{lem}
Proof is by direct but rather cumbersome calculation. \par
Thus, one has
\begin{prop}
 Closedness of $2g+1$ one-forms in $W_{gc}$ is the necessary and sufficient condition for closedness of the one-form
\begin{equation}
\label{w-wk-S2n}
 \omega(z)=\sum_{k=1-2g}^\infty z^{-k} \omega_k =\sum_{i=0}^{2g} p_{2(g+i)+1}(z)\ dx_{2(g+i)+1}. 
\end{equation}
\end{prop}
 \begin{cor}
  For the variety $W_{gc}^I$  one has locally
\begin{equation}
  \label{wdS-S2n}
\omega(z)=dS(z,x)
\end{equation}
where 
\begin{equation}
\label{S-Sn}
 S(x,z)=\sum_{k=g}^{3g} z^{2k+1}x_{2k+1}+\sum_{l=g}^1z^{2l-1}S_{1-2l}+\sum_{m=0}^\infty \frac{S_{2m+1}}{z^{2m+1}}
\end{equation}
 \end{cor}
and
\begin{equation}
\label{P-H-S-S2n}
 \begin{split}
  P_{2j+1}(z)=&\frac{\partial S(x,z)}{\partial x_{2j+1}}, \qquad j=-g,\dots,3g \\
H^{2j+1}_k= &\frac{\partial S_k}{\partial x_{2j+1}}, \qquad k=1-2g, 3-2g, 5-2g \dots\ .
\end{split}
\end{equation}
In virtue of (\ref{P-H-S-S2n}) the algebraic relations (\ref{Halgfin-S2n}), (\ref{H-P-S2n}) becomes systems of $2g+1$ PDEs of Hamilton-Jacobi type for $2g+1$ unknown $S_{1-2g}$, $S_{3-2g}$, $\dots$, $S_1$, $\dots$, $S_{2g+1}$. In particular, the system (\ref{Halgfin-S2n}) takes the form
\begin{equation}
\label{DS-Halgfin-S2n}
 \begin{split}
 \frac{\partial S_{1-2k}}{\partial x_{2g+3}} - \frac{\partial S_{3-2k}}{\partial x_{2g+1}}
+\frac{\partial S_{2g-1}}{\partial x_{2g+1}}\frac{\partial S_{2k-1}}{\partial x_{2g+1}}= & 0, \qquad k=1-g,2-g,\dots,g \\
 \frac{\partial S_{1+2g}}{\partial x_{2g+3}}-
\frac{\partial S_{2g+3}}{\partial x_{2g+1}}\left(\{\frac{\partial S_{1-2l}}{\partial x_{2g+1}}\}_{l=-g,\dots,g} \right)+ 
  \frac{\partial S_{2g-1}}{\partial x_{2g+1}} \frac{\partial S_{2g+1}}{\partial x_{2g+1}} =&0 , 
 \end{split}
\end{equation}
while the relations (\ref{H-P-S2n}) give
\begin{equation}
\label{DS-H-P-S2n}
 \frac{\partial S_{1-2n}}{\partial x_{2(g+k)+1}}=
F_{nk}\left(\{\frac{\partial S_{1-2l}}{\partial x_{2g+1}}\}_{l=-g,\dots,g} \right) \qquad n=-g,\dots,g,\ k=2,3,4,\dots,2g
\end{equation}
where $F_{nk}$ are certain polynomials on $\frac{\partial S_{1-2l}}{\partial x_{2g+1}}$.\par
It is a straightforward check that the systems of PDEs (\ref{DS-Halgfin-S2n}), (\ref{DS-H-P-S2n}) commute between themselves. So one can state the following
\begin{prop}
 Variety $W_g^I$ is a family of deformations of the coordinate rings of the hyperelliptic curves (\ref{C2g+1}) governed by the $2g$ commuting $2g+1$ component systems of PDE (\ref{DS-Halgfin-S2n}), (\ref{DS-H-P-S2n}).
\end{prop}
As the concrete example we take the relations (\ref{H7i-S4}) for $g=2$. The corresponding system of PDEs is
 \begin{equation}
\label{DS-H7i-S4} 
\begin{split}
\frac{\partial S_{-3}}{\partial x_7}=& \frac{\partial S_{-1}}{\partial x_5}
-\left(\frac{\partial S_{-3}}{\partial x_5}\right)^{2},\\
\frac{\partial S_{-1}}{\partial x_7}=& \frac{\partial S_{1}}{\partial x_5}-
\frac{\partial S_{-3}}{\partial x_5}\frac{\partial S_{-1}}{\partial x_5},\\
\frac{\partial S_{1}}{\partial x_7}=& \frac{\partial S_{3}}{\partial x_5}-
\frac{\partial S_{-3}}{\partial x_5}\frac{\partial S_{1}}{\partial x_5},\\
\frac{\partial S_{3}}{\partial x_7}=&  \frac{\partial S_{5}}{\partial x_5}-
\frac{\partial S_{-3}}{\partial x_5}\frac{\partial S_{3}}{\partial x_5},\\
\frac{\partial S_{5}}{\partial x_7}=&  
-\frac{\partial S_{-1}}{\partial x_5}\frac{\partial S_{3}}{\partial x_5}
-\frac{1}{2}\,\left(\frac{\partial S_{1}}{\partial x_5}\right)^{2}
-2\frac{\partial S_{-3}}{\partial x_5}\frac{\partial S_{5}}{\partial x_5}.
 \end{split}
\end{equation}
The relation (\ref{H9i-S4}) and those for $H^{11}_i,H^{13}_i$ give rise to the three other systems of PDEs of Hamilton-Jacobi type. These four 5 components systems of PDEs describe special class of deformations of the $g=2$ curve (\ref{C2g+1}) parameterized by $5$ variables. \par
The systems (\ref{DS-Halgfin-S2n}), (\ref{DS-H-P-S2n}) have various equivalent forms. For example, introducing $v_j=\frac{\partial S_{j}}{\partial x_{2g+1}}$, $j=1-2g,\dots,1+2g$, one rewrites the system (\ref{Halgfin-S2n}), (\ref{H-P-S2n}) as the set of $2g$ systems of conservation laws
\begin{equation}
 \label{cons-law-Sn}
\frac{\partial v_{2j+1}}{\partial x_{2(g+k)+1}}=\frac{\partial}{\partial x_{2g+1}} F_{2j+1,k}(v), \qquad j=-g,\dots,g,\ k=1,2,\dots,2g.
\end{equation}
 In terms of the coefficients $u_j$ of the hyperelliptic curve (\ref{C2g+1}) the system (\ref{Halgfin-S2n}), (\ref{H-P-S2n}) become the set of systems of hydrodynamical type
\begin{equation}
 \label{Mat-V-coeff-Sn}
\frac{\partial u_{2j+1}}{\partial x_{2(g+k)+1}}= V^{(k)}_{jl}(u) 
\frac{\partial u_l}{\partial x_{2g+1}}  , \qquad j=0,1,2,\dots,2g 
\end{equation}
where $V^{(k)}_{jl}(u)$ are certain polynomials on $u_j$.\par
For example, the system (\ref{DS-H7i-S4}) takes the forms
\begin{equation}
\label{cons-law-S4}
 \begin{split}
 \frac{\partial v_{-3}}{\partial x_7}=& \frac{\partial}{\partial x_5} \left(v_{-1}-(v_{-3})^2 \right),\\
 \frac{\partial v_{-1}}{\partial x_7}=& \frac{\partial}{\partial x_5} \left(v_{1}-v_{-3}v_{-1} \right),\\
 \frac{\partial v_{1}}{\partial x_7}=& \frac{\partial}{\partial x_5} \left(v_{3}-v_{-3}v_{1} \right),\\
 \frac{\partial v_{3}}{\partial x_7}=& \frac{\partial}{\partial x_5} \left(v_{5}-v_{-3}v_{3} \right),\\
 \frac{\partial v_{5}}{\partial x_7}=& \frac{\partial}{\partial x_5} \left(-2v_{-3}v_{5} -v_{-1}v_{-2}-\frac{1}{2}(v_{1})^2 \right),
 \end{split}
\end{equation}
and
\begin{equation}
 \label{Mat-V-coeff-S4}
\begin{split}
 \frac{\partial u_{4}}{\partial x_7}=& {\frac {\partial }{\partial x_{{5}}}}u_{{3}}  
-\frac{3}{2}\, \left( {\frac {\partial }{\partial x_{{5}}}}u_{{4}} 
  \right) u_{{4}},  \\ 
 \frac{\partial u_{3}}{\partial x_7}=& {\frac {\partial }{\partial x_{{5}}}}u_{{2}}  
- \left( {\frac {\partial }{\partial x_{{5}}}}u_{{4}}   \right) 
u_{{3}}  -\frac{1}{2}\, \left( {\frac {\partial }{\partial x_{{5}}}}u_{{3}}  
 \right) u_{{4}},  \\
\frac{\partial u_{2}}{\partial x_7}=& {\frac {\partial }{\partial x_{{5}}}}u_{{1}}
- \left( {\frac {\partial }{\partial x_{{5}}}}u_{{4}}   \right) u_{{2}}  
-\frac{1}{2}\, \left( {\frac {\partial }{\partial x_{{5}}}}u_{{2}}  
 \right) u_{{4}},    \\
\frac{\partial u_{1}}{\partial x_7}=& {\frac {\partial }{\partial x_{{5}}}}u_{{0}}  
-\frac{1}{2}\, \left( {\frac {\partial }{\partial x_{{5}}}}
u_{{1}}   \right) u_{{4}}  
- \left( {\frac {\partial }{\partial x_{{5}}}}u_{{4}}  
 \right) u_{{1}},  \\
\frac{\partial u_{0}}{\partial x_7}=&  -\frac{1}{2}\, \left( {\frac {\partial }{\partial x_{{5}}}}u_{{0}} 
  \right) u_{{4}}  - 
\left( {\frac {\partial }{\partial x_{{5}}}}u_{{4}}  \right) u_{{0}}.  
\end{split}
\end{equation}
The systems(\ref{DS-Halgfin-S2n}), (\ref{DS-H-P-S2n}), (\ref{cons-law-Sn}) and (\ref{Mat-V-coeff-Sn}) represent three different forms of the same system of flows which describe deformations of the hyperelliptic curve (\ref{C2g+1}).\par
For the whole variety $W_g$ a special subvariety $W^I_g$ is defined by the requirement that  $2g+1$ one-forms
\begin{equation}
 \omega_i =\sum_{k=0}^\infty H^{2(g+k)+1}_{2i+1} dx_{2(g+k)+1},\qquad i=-g,\dots,g\ ,
\end{equation}
 where $x_{2(g+k)+1}$, $k=0,2,\dots$ are local coordinates in $W_g$, are closed. This condition is equivalent to the condition of closedness of the one-form
\begin{equation}
\label{w-Sn}
 \omega(z) =\sum_{k=0}^\infty p_{2(g+k)+1}(z) dx_{2(g+k)+1}.
\end{equation}
The system (\ref{Mat-V-coeff-S4}) and the corresponding system for $g>2$ are the well known examples  
of the integrable hydrodynamical type systems, called the dispersionless coupled KdV systems. They have all properties typical for integrable systems: infinite sets of symmetries, conservation laws, bi-Hamiltonian structures (see e.g. \cite{FP,KK,KMAM-10}). In the paper \cite{KK} they arose as the hidden BH hierarchies in the Birkhoff strata and the compact form of such hierarchies has been found too. Also the fact that they can be written in the form (\ref{cons-law-Sn}) of conservation  laws also follows from their representation found in that paper.\par
 We would like to emphasize that in our approach they arise in a manner which is completely different from the previous one: they describe local properties of a special class of algebraic varieties in $W^I_g$ in the Birkhoff strata Gr$^{(2)}$. \par 
One more feature of the approach presented here is that it reveals a close interrelation between the special algebraic varieties of the type (\ref{H2n-S2g+1}), (\ref{H2n+1-S2g+1}) and integrable hydrodynamical type systems (\ref{Mat-V-coeff-Sn}).\par
Similar connection in different setting, namely, between hyperbolic systems of conservation laws and congruences of lines in projective spaces has been noted and studied in the papers \cite{AF1,AF2,AF3}. Comparison of the formulae (\ref{Halgfin-S2n}), (\ref{H-P-S2n}) and (\ref{cons-law-Sn}) with the first formulae from the papers \cite{AF1,AF2,AF3} indicates that these two approaches could be connected. However one can show that, for instance, the system (\ref{Eqn-ui5-S1}) does not belong to the Temple class studied in \cite{Temple}.  
\section{Ideals of varieties $\mathbf{W^I_g}$ as Poisson ideals}
Any cotangent bundle carries a natural symplectic structure (see e.g. \cite{Sha,Arn}). Formulae (\ref{w-wk-S2n}), (\ref{wdS-S2n}) indicate that the symplectic structure  on $T^*_{W_{gc}}$ should be
\begin{equation}
 \label{Omegagc-S2n}
\Omega_{gc}=\sum_{k=0}^{2g} dp_{2(g+k)+1} \wedge dx_{2(g+k)+1}.
\end{equation}
In virtue of (\ref{wdS-S2n}) for the subvariety $W_{gc}^I$ one has
\begin{equation}
\Omega_{gc}\big{|}_{W_{gc}^I}=0,
\end{equation}
i.e. the subvariety $W_{gc}^I$ is the Lagrangian subvariety in $W_{gc}$ if one treats $W_{gc}$ as a symplectic variety equipped with the symplectic two-form (\ref{Omegagc-S2n}) and $p_j,x_j$ being the classical Darboux coordinates. \par
Passing then to the infinite dimensional varieties $W_g$, i.e. the $2g$-parametric families of the coordinate rings for the hyperelliptic curve, one should consider an infinite-dimensional symplectic varieties equipped with 2-form
\begin{equation}
 \label{Omegag-S2n}
\Omega_{g}=\sum_{k=0}^{\infty} dp_{2(g+k)+1} \wedge dx_{2(g+k)+1}.
\end{equation}
 Within such symplectic interpretation it is quite natural to require that ideal $I(W_g)$ of the variety $W_g$ has some properties typical for symplectic or Poisson  varieties. One of the most natural requirement is that the ideal $I(W_g)$ is a Poisson ideal, i.e.
\begin{equation}
\label{coiso}
 \left\{ I(W_g),I(W_g)\right\} \subset I(W_g)
\end{equation}
where $\{,\}$ is the Poisson bracket. The condition (\ref{coiso}) of closedness of the ideal $I$ has been used in \cite{KO-Coiso} to define the so-called coisotropic deformations of algebraic varieties. Earlier this idea has been proposed 
in \cite{KM} in the context of coisotropic deformations of commutative associative algebras.\par
The ideal of the variety $W_g$ is given by (\ref{IGamma})  with $C_{2g+1}$ and $l^{(g)}_{2k+1}$, defined by (\ref{C2g+1}) and (\ref{lg}) or, equivalently,
\begin{equation}
 I(W_g)=\langle C_{2g+1}, \{M_k\}_g \rangle
\end{equation}
 where $M_k$ is given by the l.h.s. of (\ref{p2g+3}),i.e.
\begin{equation}
 \label{Mk}
M_k=p_{2(g+k)+3}-\lambda p_{2(g+k)+1}+H^{2(g+k)+1}_{-(2g-1)}p_{2g+1}, \qquad k=g,g+1,g+2,\dots\ .
\end{equation}
In basis of the ideal $I(W_g)$ composed by $C_{2g+1}$ and $M_k$ the closedness condition (\ref{coiso}) with canonical Poisson bracket for Darboux coordinates $x_j,p_j$ is equivalent to the following
\begin{equation}
 \label{coiso-CM}
\begin{split}
 \{C_{2g+1},M_k\}=&-\frac{\partial H^{2(g+k)+1_{1-2g}}}{\partial x_{2g+1}} C_{2g+1}, \\
\{M_l,M_k\}=& 0,
\end{split}
\end{equation}
while $H^j_k$ and $u_m$ should obey the differential equations
\begin{equation}
  \frac{\partial H^{2(g+k)+1_{1-2g}}}{\partial x_{2(g+l)+1}}=
 \frac{\partial H^{2(g+l)+1_{1-2g}}}{\partial x_{2(g+k)+1}}, \label{dHdH}\\ 
\end{equation}
\begin{equation}
\frac{\partial H^{2(g+k)+1_{1-2g}}}{\partial x_{2(g+l)+3}}
\frac{\partial H^{2(g+l)+1_{1-2g}}}{\partial x_{2(g+k)+3}}
+H^{2(g+l)+1}_{1-2g}\frac{\partial H^{2(g+k)+1_{1-2g}}}{\partial x_{2g+1}}
-H^{2(g+k)+1}_{1-2g}\frac{\partial H^{2(g+l)+1_{1-2g}}}{\partial x_{2g+1}}=0, \label{dHdH2}\\
\end{equation}
\begin{eqnarray}
&& \frac{\partial u_m}{\partial x_{2(g+k)+3}}
-(1-\delta_{m,0})\frac{\partial u_{m-1}}{\partial x_{2(g+k)+1}}
-H^{2(g+k)+1}_{1-2g}\frac{\partial u_m}{\partial x_{2g+1}}
-2u_m\frac{\partial H^{2(g+k)+1_{1-2g}}}{\partial x_{2g+1}}
=0, \label{dudH}\\
&& k,l=0,1,2,\dots,\qquad m=0,1,2,\dots,2g\ ,  \nonumber
\end{eqnarray}
where $\delta_{m,0}$ is the Kronecker symbol. This is an infinite hierarchy of equations for $2g+1$ unknowns $H^{2g+1}_{1-2g}$, $H^{2g+1}_{3-2g}$, \dots, $H^{2g+1}_{1}$, $H^{2g+1}_{2g+1}$, since all  $H^{2(g+k)+1}_{1-2g}$ are polynomials of these $2g+1$ variables.\par
It is a straightforward check that the system (\ref{dHdH})-(\ref{dudH}) is equivalent to the hierarchy of the systems associated with the system (\ref{Mat-V-coeff-Sn}). In other words the hydrodynamical type systems discussed in the previous sections represent coisotropic deformations of curves (\ref{C2g+1}). We note that in our approach these systems arise within the study of local properties of special subvariety $W_{gc}$ carried a set of $2g+1$ closed 1-forms.\par
Within the symplectic interpretation of the varieties $W_g$ one has simple realizations of the 2-cocycles namely
\begin{equation}
\label{2-cocy-Poiss}
 \psi_g(p_j,p_k)=\{\alpha,f_{jk}\}|_{W_g}
\end{equation}
 where $\alpha=\sum_{i \geq 2g+1} \alpha_i p_i$, $\alpha_i$ are arbitrary constants.
\section{Cohomology blow-ups and gradient catastrophe}
In the previous sections it was shown that dispersionless coupled KdV (dcKdV) hierarchies (BH hierarchy for $g=0$) provide us with a special class of 2-cocycles and 2-coboundaries which are well defined for regular solutions of the dcKdV hierarchies. Singular solutions of dcKdV of these hierarchies give rise to a singular behavior of 2-cocycles and 2-coboundaries.\par
It is well known (see e.g. \cite{DN}) that the singular sector for the BH equation and BH hierarchy is composed by solutions which exhibit gradient catastrophe when derivatives of their solutions blow-up (become unbounded). Similar situation takes place also for dcKdV hierarchies \cite{KK}. Formulae (\ref{2-cocy-S0}), (\ref{2-cob-S0}), (\ref{2cocy-S1}), (\ref{2-cocy-Poiss}) readily show that blow-ups of $\frac{\partial u_k}{\partial x_l}$  lead to blow-up of the corresponding 2-cocycles and 2-coboundaries (unboundedness of their values). Thus gradient catastrophe for BH and dcKdV hierarchies and blow-ups for Harrison cohomology of the subvarieties $W^I_g$ are intimately connected. Analysis of the singular sectors of the BH hierarchy and dcKdV hierarchies performed in \cite{KK,KMAM-10} show that the gradient catastrophe for these hierarchies happens on the subvarieties in the space of independent variables $x_j$ of finite codimensions and it is associated with the transition from one Birkhoff stratum to another. So, such a transition is accompanied also by the blow-up of 2-cocycles and 2-coboundaries. Whether or not the blow-ups of Harrison cohomology happen on the subspaces of finite codimension (i.e. of zero measure) and are associated with certain  transition between different strata is an open problem. 
\section{Discussion}
Theory of finite-gap solutions and the theory of Whitham equations are two most known chapters in the story of interconnection between algebraic curves and integrable nonlinear PDEs. Theory  of periodic (or finite-gap) solutions for the KdV equation has been first discussed in the papers \cite{Nov2,Lax,Dub2,IM,mKvM} and then has been extended to many other integrable nonlinear PDEs (see e.g. \cite{DMN,Kri1,Kri2,KN2,Dub3,Kri4,Mul2}, review \cite{Mat2} and references therein). In this theory, an algebraic curve (hyperelliptic curve in most cases) is fixed by initial data and, hence, remains unchanged during the evolution. Choosing different curves, e.g. hyperelliptic curves of different genus, one constructs classes of exact solutions of the same integrable PDE \cite{Nov2,Mat2}.\par
Situation is quite different in Whitham theory. Whitham was the first to observe \cite{Whi1} (see also \cite{Whi2})   
that the description of slow modulations of the simple phase travelling wave solutions of the KdV equation requires to consider an elliptic curve with varying parameters (moduli). Within the averaging method developed in \cite{Whi1,Whi2} the variations of the corresponding elliptic curve
\begin{equation}
\label{f-ell}
 q^2=\left(\mu - \beta_1\right)\left(\mu - \beta_2\right)\left(\mu - \beta_3\right)
\end{equation}
are described by the system of equations (see e.g. \cite{ZMNP})
\begin{equation}
\label{Whi-eqn}
 \frac{1}{4}\frac{\partial \beta_i}{\partial T}+ v_i(\beta_1,\beta_2,\beta_3) \frac{\partial \beta_i}{\partial X}=0, \qquad i=1,2,3 
\end{equation}
where $X,T$ are slow variables and the characteristic velocities $v_i$ are
\begin{equation}
\label{Whi-eqn2}
 \begin{split}
  v_1=& \frac{1}{2}(\beta_1+\beta_2+\beta_3)+(\beta_1-\beta_2) \frac{K(s)}{E(s)}, \\
  v_2=& \frac{1}{2}(\beta_1+\beta_2+\beta_3)+(\beta_2-\beta_1) \frac{sK(s)}{E(s)-(1-s)K(s)}, \\
  v_3=& \frac{1}{2}(\beta_1+\beta_2+\beta_3)+(\beta_2-\beta_3) \frac{K(s)}{E(s)-K(s)}.
 \end{split}
\end{equation}
Here $s=\frac{\beta_2-\beta_3}{\beta_1-\beta_3}$ and $K(s)$ and $E(s)$ are complete elliptic integrals of the first and second kind.\par
Reformulation of Whitham equations (\ref{Whi-eqn}), (\ref{Whi-eqn2}) in terms of abelian differentials on a Riemann surface and generalization to modulated N-phase KdV waves associated with hyperelliptic curves given in  \cite{FFM,LL}
clearly demonstrated that ``\dots the description of the modulations of finite-gap solutions of KdV requires the introduction of a {\it family of Riemann surfaces varying gradually with X and T}. The mathematical problem is therefore one of {\it deformation} of hyperelliptic curves'' (\cite{FFM} p. 740). Further works on Whitham equations
(see e.g. \cite{Kri5,DN,Dub4,Kri6,FRT,Kri3,KMWZ}) provide us with a wide class of deformations of algebraic curves, in particular, of hyperelliptic ones. \par
The formulation of deformations of hyperelliptic curves considered in the present paper looks, at first sight, similar to that from Whitham theory, particularly to that of Krichever's universal Whitham hierarchy \cite{Kri6}. In fact, some objects like the action $S(z)$ (\ref{S-S0},\ref{S-Sn})  and closed one-forms (\ref{w-S0},\ref{w-Sn}) are common in both approaches. However, the comparison of the Whitham deformations of hyperelliptic curves
 and those given in sections 5--8  as well as in the papers \cite{FP,KK,KMAM-10,KO-Coiso} clearly shows their difference. Let us consider the simplest case of an elliptic curve. Whitham deformations of the curve 
(\ref{f-ell}) are described by the system of equations (\ref{Whi-eqn}), (\ref{Whi-eqn2}). On the other hand equations  
 (\ref{Eqn-ui5-S1}) which describe our deformation of the curve 
\begin{equation}
\label{f-ell-ns}
 p^2=\left(\lambda - \gamma_1\right)\left(\lambda - \gamma_2\right)\left(\lambda - \gamma_3\right) 
\end{equation}
rewritten in terms of the Riemann invariants $\gamma_1,\gamma_2,\gamma_3$ are of the form \cite{FP}  
\begin{equation}
\label{eb1-S1}
\frac{\partial \gamma_i}{\partial x_5}= \frac{1}{2}(\gamma_1+\gamma_2+\gamma_3+2 \gamma_i) \frac{\partial \gamma_i}{\partial x_3}, \qquad i=1,2,3.
\end{equation}
The systems (\ref{Whi-eqn}), (\ref{Whi-eqn2}) and (\ref{eb1-S1}) are apparently quite different. 
Moreover, it is easy to see that there is no a birational transformation $(p,\lambda) \to (q,\mu)$ between the elliptic curves (\ref{f-ell}) and (\ref{f-ell-ns}) which simultaneously converts the system (\ref{eb1-S1}) into the system (\ref{Whi-eqn}), (\ref{Whi-eqn2}).\par
We would like to emphasize that in our approach algebraic curves are completely defined by the structures of the Birkhoff strata of the Grassmannian Gr$^{(2)}$ and integrable systems of hydrodynamical type arise and are associated with the special subvarieties in Birkhoff strata. 
Possible interrelation between Whitham deformations and coisotropic deformations of hyperelliptic curves and other algebraic curves is an intriguing open problem. We hope to clarify it in future publication. \par
Cohomological structure of algebraic varieties and integrable equations in various settings have been discussed earlier e.g. in \cite{Mul4,NS,Nak1,CK}. Harrison cohomology for the families of the Veronese, elliptic and hyperelliptic curves studied in the present paper seems to be quite different from those considered before. Possible interconnection between all these cohomological constructions will be considered elsewhere. 
\subsubsection*{Acknowledgements} The authors are grateful to E. Ferapontov  for numerous fruitful discussions. The authors also thank one of the Referees for careful reading of the manuscript and several useful comments. This work has been partially supported by PRIN grant no 28002K9KXZ and by FAR 2009 (\emph{Sistemi dinamici Integrabili e Interazioni fra campi e particelle}) of the University of Milano Bicocca.

\end{document}